\documentclass[twocolumn]{aastex631}
\usepackage{amsmath}
\usepackage{booktabs} 
\usepackage{savesym}
\savesymbol{tablenum}
\usepackage{siunitx}
\restoresymbol{SIX}{tablenum}

\usepackage{txfonts}
\usepackage{color}
\usepackage{mathtools}
\usepackage{CJK}
\usepackage[mathlines]{lineno} 

\newcommand{\actaken}[2]{{#1}}
\newcommand{\grtaken}[2]{{#1}}

\newcommand{\EY}{\mathrm{EY}}

\newcommand{\m}{{\rm ~m}}
\newcommand{\km}{{\rm ~km}}

\newcommand{\s}{{\rm ~s}}
\newcommand{\h}{{\rm ~h}}

\newcommand{\J}{{\rm ~J}}
\newcommand{\K}{{\rm ~K}}

\newcommand{\B}{{\rm B}}

\newcommand{\Y}{{\rm Y}}

\newcommand{\p}{{\rm p}}

\shorttitle{The EY effect in planetary rings}
\shortauthors{Wen-Han Zhou et al}

\begin{document}


\title{Dynamics of planetary rings under thermal forces}

\correspondingauthor{Wen-Han Zhou}
\email{wenhan.zhou@oca.eu}

\author[0000-0003-4229-8936]{Wen-Han Zhou}
\affiliation{JSPS International Research Fellow, Department of Earth and Planetary Science, The University of Tokyo, Tokyo, Japan}
\author[0000-0002-5486-7828]{Eiichiro Kokubo}
\affiliation{Division of Science, National Astronomical Observatory of Japan, Mitaka, Tokyo, Japan}
\affiliation{Center for Computational Astrophysics, National Astronomical Observatory of Japan, Tokyo, Japan}
\affiliation{Department of Astronomy, University of Tokyo, Tokyo, Japan}
\author[0000-0002-3544-298X]{Harrison Agrusa}
\affiliation{Universit\'e C\^ote d'Azur, Observatoire de la C\^ote d'Azur, CNRS, Laboratoire Lagrange, Nice, France}
\affiliation{Centre national d'\'etudes spatiales (CNES), Paris, France}
\author[0009-0002-5271-8658]{Gregorio Ricerchi}
\affiliation{Universit\'e C\^ote d'Azur, Observatoire de la C\^ote d'Azur, CNRS, Laboratoire Lagrange, Nice, France}
\author[0000-0002-1293-9782]{Aur\'elien Crida}
\affiliation{Universit\'e C\^ote d'Azur, Observatoire de la C\^ote d'Azur, CNRS, Laboratoire Lagrange, Nice, France}
\author[0000-0002-6034-5452]{David Vokrouhlick\'y}
\affiliation{Astronomical Institute, Charles University, V Hole\v{s}ovi\v{c}k\'ach 2, CZ 18000, Prague 8, Czech Republic}
\author[0000-0003-4045-9046]{Yun Zhang}
\affiliation{Department of Climate and Space Sciences and Engineering, University of Michigan, Ann Arbor, MI 48109, USA}
\author[0000-0002-1772-1934]{Ronald-Louis Ballouz}
\affiliation{Johns Hopkins University Applied Physics Lab, Laurel, MD, USA}

\begin{abstract}
Planetary rings provide natural laboratories for studying the fundamental processes that govern the evolution of planetary systems. However, several key features, such as the sharp inner edges of Saturn’s rings remain unresolved. In this work, we introduce and quantify the Eclipse–Yarkovsky (EY) effect, a thermal torque arising from asymmetric thermal emission of particles during planetary eclipses, which is effective for particles larger than millimeters in size. We formulate this effect within a continuum framework appropriate for collisionally coupled planetary rings and derive the continuum evolution equation that includes the EY torque and viscous diffusion (Eq.~\ref{eq:full_evolution}), constraining its magnitude using ring particle spin distributions obtained from $N$‑body simulations. We find that the EY effect systematically produces a positive angular momentum flux that could overcome the viscous torque, driving ring material outward and leading to long‑term decretion. The total EY torque principally depends on the optical depth, in which we identify three dynamical regimes: dense, transitional, and tenuous regimes, each exhibiting distinct evolutionary pathways. In the dense or transition regimes, the EY torque can produce a sharp inner edge such as that of Saturn’s A ring. In the tenuous regime, it can drive an entire ring outward while preserving shape. This outward transport may also facilitate satellite formation beyond the Roche limit. {We also quantitatively show that planetary thermal radiation on rings exerts an opposing torque, namely planetary-Yarkovsky effect, whose importance depends on planetary emissivity and ring-particle albedo, and may lead to inward transport in Saturn's close-in rings.}
\end{abstract}





\section{Introduction} \label{sec:intro}

\begin{figure*}
    \centering
    \includegraphics[width= \linewidth]{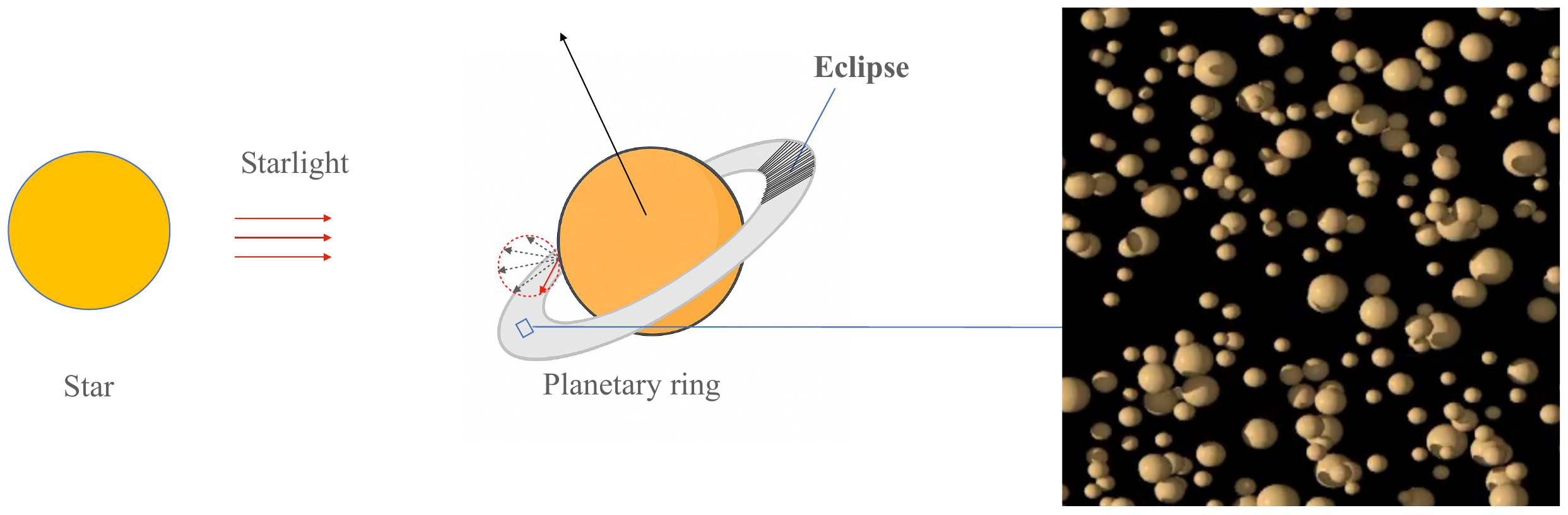}
    \caption{Schematic of the planetary ring with a shadow. The planet is surrounded by a ring with a tilted angle $\sim 20^\circ$, {mirroring the configurations of Mars or Saturn}. {The radiation from a surface element of the planet follows Lambert emission low, denoted by the arrows within the red dashed circle.} The thermal force arises from the thermal emission of particles. In particular, the eclipse leads to an asymmetric thermal emission of particles and therefore produces a net thermal torque after averaging the orbit, namely the Eclipse-Yarkovsky effect. The right panel shows the simulated box space with ring particles as an example.}
    \label{fig:schematic}
\end{figure*}

All giant planets in the Solar System host rings composed of small particles (typically $\leq$ a few meters), with Saturn’s rings being the most massive and well studied. These systems serve as natural laboratories for exploring key dynamical processes—collisions, self-gravity, resonances, and angular momentum transport—that govern planetary system evolution \citep[e.g.][]{Goldreich1978a, Goldreich1978b, Fuller2014,Esposito2014, Tiscareno2018b, Torri2025, Kambara2025}. Over time, viscous spreading can drive material beyond the Roche limit to form moons, explaining the observed mass–distance distribution of planetary satellites \citep{Crida2012}. Mars’ moons Phobos and Deimos have similarly been proposed to originate from a past ring system \citep{Hesselbrock2017}.

However, several puzzles remain in our understanding of ring evolution and structure. One long-standing issue is the presence of sharp edges in Saturn’s major rings. While the outer edges can be explained by the balance between positive viscous torques and negative Lindblad torques from nearby moonlets, the sharp inner edges (e.g., of the A and B rings) are not well understood due to the lack of confining inner satellites that could supply the necessary resonant torques \citep[e.g.][]{Crida2025}. One proposed mechanism is ballistic transport, in which high-speed impacts from micrometeoroids eject debris from ring particles \citep{Ip1984, Cuzzi1990, Durisen1984, Durisen1989}. These ejecta follow ballistic trajectories and re-accrete elsewhere in the ring, effectively redistributing mass and angular momentum. Ballistic transport tends to move the mass and angular momentum of the inner edge outwards and is capable of maintaining sharp edges. This process cannot create sharp edges; it can only maintain an existing sharp edge. Another puzzle is related to a theoretical past Martian ring which would have existed if Phobos and Deimos formed from a giant impact followed by accretion from a circumplanetary disk \citep{Singer1967, Strom1992, Craddock2011, Canup2018, Citron2015, Rosenblatt2016}. Dynamical models suggesting that past Martian rings should persist for billions of years due to the long timescale of viscous spreading \citep{Hesselbrock2017, Madeira2023}, but no ring is observed today. Together, these issues point to missing physics in current ring evolution models.

In this work, we introduce a previously underappreciated thermal recoil effect, referred to here as the Eclipse–Yarkovsky (EY) effect, as a mechanism that may help resolve these puzzles. While this paper mostly takes Saturn's ring systems as the example, the theory developed is sufficiently general so that it is applicable to other planetary ring systems.

This paper is organized as follows: Sec.~\ref{sec:overview} offers an overview and historical context of the EY effect. Sec.~\ref{sec:theory} derives the general continuum equation form with the EY term given a particle size-frequency distribution (SFD). Section \ref{sec:numerical} calculates the EY term based on the particle rotation distribution obtained from the $N$-body simulations, and Sec.~\ref{sec:evolution} presents the long-term evolution by numerically solving the continuum equation incorporating the EY effect and discusses different regimes of the EY effect according to the optical depth. Finally we discuss the implications of the EY effect on the ring systems in the Solar system in Sec.~\ref{sec:discussion} and summarize our findings in Sec.~\ref{sec:conclusion}.

\section{The Eclipse-Yarkovsky effect: an overview and historical context}
\label{sec:overview}

\subsection{Radiation forces in celestial mechanics: historical background}
\label{subsec:history}

{The concept of radiation pressure in modern physics arouse from the development of Maxwell's electromagnetic theory by a group of enthusiastic students and followers after his premature death in 1879, with Poynting playing a central role in clarifying its physical implications \citep[see][for a concise overview]{lb2012}. Although there existed earlier speculations about the role of sunlight pressure in the context of astronomical phenomena earlier, these lacked both quantitative theoretical estimates and experimental confirmation. This changed in the early twentieth century, when \citet{lebedew1910} experimentally confirmed the existence of radiation pressure in agreement with Maxwell’s theory.}

{Heliocentric orbital dynamics are directly influenced by such radiation-driven forces. While the direct sunlight pressure is conservative, other radiation-related non-conservative effects were also discovered and accurately described. Given today's perspective, the prime examples consist of (1) the Poynting-Robertson effect \citet{ww1950} \citep[based on its original formulation by][]{p1904,r1937}, and (2) the Yarkovsky effect resurrected by \citet{o1951}. \citet{bls1979} remains a masterful overview of the radiation processes and their role in planetary astronomy in the second half of the 20th century, with additional information on the Yarkovsky effect in \citet{Bottke06} and \citet{Vokrouhlicky2015}.} 

{The essence of the Yarkovsky effect consists of the recoil due to the thermal radiation of the body. Because of thermal inertia, the emission over the surface is anisotropic: the hotter afternoon side experiences larger recoil than the cooler night side of the surface. The net effect integrated over the whole surface is generally tilted from the radial direction with a non-zero along-track projection. As a result, the effect leads to an accumulated change in semimajor axis, either positive or negative
depending on the rotational state. This is the solar diurnal variant of the effect, conceptually pictured by \citet{o1951} and summarized in \citet{bls1979} (with more intermediate reference there in). In the next decades, it was realized that there also exists a solar seasonal \footnote{In this work, we define ''solar season'' as the (thermal) variations driven by the planet's heliocentric orbit, while ''planetary season'' denotes the variations resulting from the rapid planetocentric orbital cycles of particles.} variant of the Yarkovsky effect \citep[][]{Rubincam1987,rubincam1988,slabinski1997}, which is caused by the thermal lag over the heliocentric orbital motion \citep[see Fig. 1 in][]{Bottke06}. The mathematical solution of the Yarkovsky effect for a spherical object was finally given by \citet{Vokrouhlicky1999}.}

{The quantitative study of thermal recoil forces was catalyzed by Earth-heated satellite dynamics \citep[e.g.,][]{Rubincam1982,Afonso1989,Rubincam1990,Farinella1990,Farinella1996,slabinski1997} and later asteroid dynamics \citep[e.g.,][]{Afonso1995, Rubincam1995, Farinella1998}. The widespread acceptance of the Yarkovsky effect by the community followed especially after the first direct detection of the heliocentric orbit drift of asteroid 6489 Golevka by \citet{chesley2003}. These studies demonstrated that even minute radiation-driven accelerations can cause significant orbital evolution over long timescales. Since then, Yarkovsky-induced transverse accelerations have been measured for hundreds of asteroids \citep[e.g., ][]{Liu2023, Fenucci2024} and are now archived online. In the asteroid population, the Yarkovsky effect is a key mechanism driving long-term orbital evolution of meter– to kilometer–scale bodies. It produces systematic semimajor-axis drift that produce near-Earth object populations from the main belt, shapes the age and structure of asteroid families, and enables constraints of asteroid bulk density and thermal properties when combined with astrometric observations \citep[see review in][]{Bottke06, Vokrouhlicky2015}.}

{Planetocentric orbital dynamics under an external heating source, which describes a situation when sunlight heats a small satellite (or dust particle) orbiting a planet, stemmed again from the field of satellite geodesy. \citet{bls1979} discuss the case of the solar radiation pressure for such a configuration, noting that the principal long-term perturbation concerns the planetocentric eccentricity and pericenter, which helps in understanding structure of Jupiter's tiny dust rings \citep[e.g.,][]{hk1996} or rarefied regions in Saturn's system \citep[e.g.,][]{hedman2013}. However, the role of planetary shadow is neglected in these early studies.}

\subsection{Eclipse-modulated Yarkovsky forces}
\label{subsec:ey_physics}

In a simple planetocentric configuration without eclipses, the solar-driven
Yarkovsky force acting on a particle typically averages to zero over one
planetocentric orbit: phases of positive and negative along-track acceleration cancel due to symmetry. This cancellation, however, is fundamentally broken when the particle experiences eclipses by the planet.

When a particle enters the planetary shadow, its surface cools and thermal
emission is temporarily suppressed. Upon re-emergence into sunlight, the surface re-heats with a delay set by the particle’s thermal inertia. The resulting asymmetry between cooling and heating phases produces a net thermal recoil force that does not vanish upon orbital averaging. Figure~\ref{fig:schematic} shows a schematic of eclipses in a planetary ring. The magnitude and sign of this effect depend on the particle’s spin state, thermal properties, and orbital geometry.

{This eclipse-modulated thermal effect has appeared in the literature under various names. It was realized in the detailed analysis of the LAGEOS spacecraft orbital decay, where a variation correlated with periods when the spacecraft orbit crossed the Earth shadow. \citet{Rubincam1982} attributed the idea to Milton Schach, coining thus the term ``Schach'' \citep{Rubincam1982} or ``Yarkovsky-Schach effect'' \citep{Rubincam1990, Farinella1996, rub2006, Vokrouhlicky2007}, even though the latter did not present any published reference. Coincidentally, the very same concept has also been previously discovered by \citet{boudon1979}, naming the effect ``photon thrust'' or ``thermal thrust'' that was used by \citet{Afonso1989} and \citet{Farinella1990}. In the context of binary asteroids \citep{Vokrouhlicky2005}, it was called the ``differential Yarkovsky'' effect in some literature \citep{Scheirich2015, Scheirich2021, Scheirich2024}. Despite the diversity of terminology, these studies describe closely related manifestations of eclipse-modulated thermal recoil arising from the same underlying physical mechanism.}

{Analytical formulations of this effect for individual objects were developed by \citet{rub2006} and \citet{Vokrouhlicky2007} and were discussed in the context of Saturn’s rings. {\citet{Rubincam2014} further considers the effect brought by the particle's thermal strain.} While early works focused on the planetary seasonal component of this effect \citep[e.g. Fig.~1 in][]{Rubincam1982}, more recent work by \citet{Zhou2024a} and \citet{Zhou2024b} revisited this effect for both planetary seasonal and diurnal components of a binary asteroid system, explicitly tracking its dependence on spin state and obliquity of asteroids. These studies showed that the diurnal component of this effect can move the object towards the synchronous rotation orbit where the spin period equals the orbital period. On top of this, for misaligned objects, the diurnal component also has components that can induce an outward drift. \citet{Zhou2024b} unified this eclipse-modulated effect with the planetary–Yarkovsky effect under the single designation of the ``binary Yarkovsky'' effect. Alongside the ``binary YORP'' effect \citep{Cuk2005}, this provides a comprehensive thermal framework for binary systems.\footnote{Unlike the binary Yarkovsky effect, binary YORP requires long-term synchronous rotation and is equally likely to generate positive or negative torques. Consequently, it does not produce a systematic effect at the level of planetary rings, where particle orientations and shapes are expected to be randomly distributed.}}

{In \citet{rub2006} and \citet{Vokrouhlicky2007}'s works in Saturn's rings, the principal caveat of these works consisted in treating the ring particles individually as if they were isolated from the surrounding population. The role of coupling between the particles was only briefly touched in Sec.~10 of \citet{rub2006} and Appendix~F of \citet{Vokrouhlicky2007}, but no analysis of the varying optical depth across the ring parts was provided. Yet, the collective phenomena were long recognized to be crucial for understanding Saturn's ring system.}

\subsection{From individual particles to a collisional continuum}
\label{subsec:continuum}

\begin{figure}
    \centering
    \includegraphics[width=\linewidth]{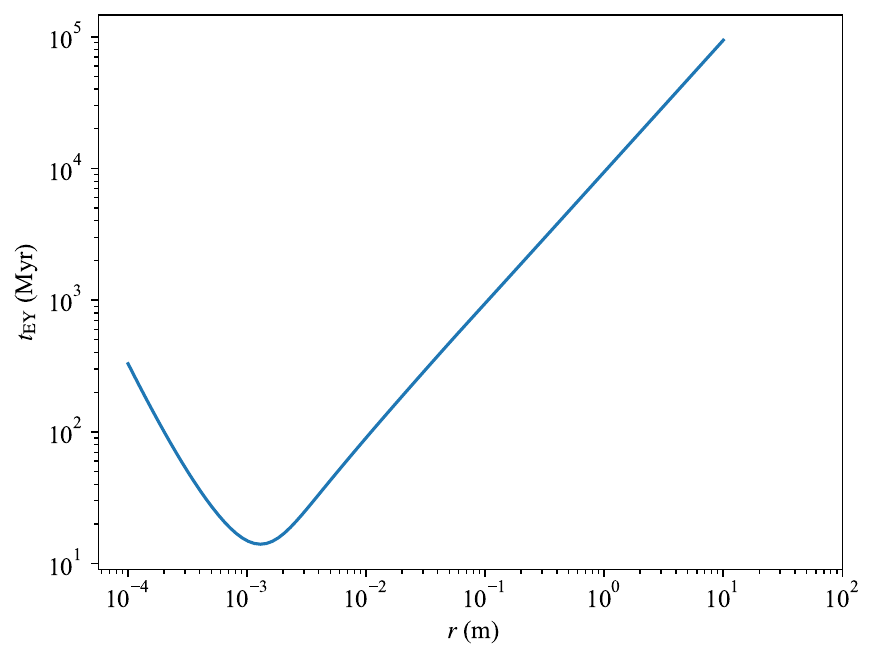}
    \caption{The timescale of the EY effect for individual particles as a function of the particle size for Saturn's system. The planetocentric distance is set to be 2 planetary radii and the EY coefficient $f_\EY = 0.002$.}
    \label{fig:timescale}
\end{figure}

{The primary objective of this work is to formulate this eclipse-modulated Yarkovsky effect within a continuum framework that would allow us to connect the potential importance of the thermal dynamical phenomena with the collective behavior of the planetary rings. The key physical insight is that, the torque (i.e., net angular-momentum flux) generated by the EY effect acting on individual particles is redistributed to neighboring particles through collisions, which conserve angular momentum. This is the essence that allows the ring annulus to respond collectively and therefore to be integrated into the continuum equation. In this setting, the relevant quantity is not the drift of an isolated object, but the redistribution of angular momentum across the ring. We therefore introduce a physically transparent term ``Eclipse--Yarkovsky effect'' that is applicable to both single-body and ring-dynamical contexts and to emphasize the central physical role of eclipses in driving the thermal recoil. At the level of individual particles, this effect reduces to the classical Yarkovsky–Schach effect (or other names in the literature) in the context of binary asteroids or Earth satellites.}

To have a preliminary insight into the importance of this effect, we estimate its timescale for individual particles \actaken{of radius $r$ orbiting at planetocentric distance $R$}{}, defined as $R/\dot R$ \citep{Zhou2024a}\actaken{:}{,}
\begin{equation}
\label{eq:t_EY}
    t_{\rm EY} \simeq 1.1~ {\rm Gyr} \left( r_\p \over 10^5 \km \right) \left( r \over 0.1 {~\rm m} \right) \left( R \over 2 ~r_{\rm p}  \right)^{1/2} \left( 0.003 \over f_{\rm EY, norm} \right) \left( a_{\rm h} \over 10{\rm ~au} \right)^2,
\end{equation}
where \actaken{}{$R$ is the planetocentric distance of the particle, }$r_{\rm p}$ is the planet's radius, $a_{\rm h}$ is heliocentric distance and {$f_{\rm EY, norm}$ is the EY coefficient anchored at $R=2~r_{\rm p}$, which measures the efficiency of the EY effect. The planet obliquity $\varepsilon_{\rm p}$ is assumed as 0 in Eq.~\ref{eq:t_EY}.} This scaling is valid only for particles larger than the thermal penetration depth $r_0 \sim 1~$mm, while a coverage over a smaller size would require the analytical solution of the EY effect (Eqs.~\ref{eq:f_EY}, \ref{eq:f_EY_d} and \ref{eq:f_EY_s}). Figure \ref{fig:timescale} shows the timescales for different particle sizes in the Saturn's ring system. We adopted $f_{\rm EY, norm} = 0.003$, as will be demonstrated in Sec.~\ref{sec:numerical}. On the other hand, the viscous evolution timescale is
\begin{equation}
    t_{\rm visc} \sim {R^2 \over \nu } \simeq 32 \,{\rm Gyrs} \left( {R \over 10^5 ~{\rm km}} \right)^2  \left( {\nu \over 10^{-2} \,{\rm m^2\, s^{-1}}}  \right)^{-1},
\end{equation}
where $\nu$ is the viscosity. The viscosity depends on the surface density and material properties, ranging from 0.003 to 0.03~$\rm m^2~s^{-1}$ for Saturn's A and B rings and being $\sim 10^{-4}$ for Saturn's C ring \citep[see e.g. ][]{Tajeddine2017, Tiscareno2018a, Crida2025}. It can be seen that the timescale for the EY effect in the Saturnian system could be shorter than the viscous evolution timescale for the size range between millimeters to meters. Due to the stronger solar radiation at Mars, the EY effect could be one hundred times stronger than for the Saturn system.

Accurate evaluation of the EY torque requires the particles’ spin-state distribution (spin rate and obliquity). Using a realistic distribution from numerical simulations, we show that the population-averaged EY torque is positive, driving particles outward (Sec.~\ref{sec:numerical}). {While the possibility of a positive eclipse-modulated torque has been noted previously \citep{rub2006, Vokrouhlicky2007},} the present work, this study offers a systematic investigation on how the EY effect affects the long-term evolution of planetary rings, by quantifying the strengths of this effect and integrating it into the continuum equations that govern ring surface density.

\section{Theory}
\label{sec:theory}

\subsection{General formula}
\label{sec:formula}

In this section, we derive the one-dimensional evolution equation for planetary rings, incorporating the angular momentum flux introduced by the Eclipse-Yarkovsky effect. The magnitude and direction of this flux will be discussed in the next section.

Consider a planetary ring characterized by a surface density profile $\Sigma(R,t)$ and a radial velocity $v_R(R,t)$ driven by dynamical mechanisms. We examine an annular segment of the ring with inner radius $R$ and radial width $\Delta R$, which therefore contains a mass of $2\pi R \Delta R \Sigma$. By mass conservation, the rate of change of mass within this annulus equals the net difference between the mass inflow through the inner boundary and the outflow through the outer boundary:
\begin{equation}
\begin{aligned}
    \frac{\partial}{\partial t} \left( 2\pi R \Delta R \Sigma \right) &= 2\pi R \cdot \Sigma(R,t) \cdot v_R(R, t) \\
    & - 2\pi (R+\Delta R) \cdot \Sigma(R+\Delta R,t) \cdot v_R(R+\Delta R, t).
\end{aligned}
\end{equation}
Taking the limit $\Delta R \to 0$, we obtain the continuity equation:
\begin{equation}
\label{eq:mass_cons}
    R \frac{\partial \Sigma}{\partial t} + \frac{\partial}{\partial R} (R \Sigma v_R) = 0.
\end{equation}
Similarly, conservation of angular momentum yields:
\begin{equation}
\begin{aligned}
\label{eq:angular_cons}
    &\frac{\partial}{\partial t} \left( 2\pi R \Delta R \cdot \Sigma R^2 \Omega \right) = 2\pi R \cdot \Sigma(R,t) R^2 \Omega(R) v_R(R,t) \\
    & - 2\pi (R+\Delta R) \cdot \Sigma(R+\Delta R,t) (R+\Delta R)^2 \Omega(R+\Delta R) v_R(R+\Delta R,t) \\
    & + \Delta J,
\end{aligned}
\end{equation}
where $\Omega(R)$ is the planetocentric mean motion frequency of the particle, and $\Delta J$ denotes the net angular momentum exchange due to additional mechanisms. A dominant contribution to $\Delta J$ arises from interparticle collisions driven by Keplerian shear (i.e., the radial differential motion between neighboring ring particles in a co-rotating Keplerian frame). This angular momentum exchange can be approximated as the difference in viscous torque across the annulus:
\begin{equation}
\begin{aligned}
    \Delta J &= -3\pi R^2 \nu(R,t) \Sigma(R,t) \Omega(R) \\
    &\quad + 3\pi (R+\Delta R)^2 \nu(R+\Delta R,t) \Sigma(R+\Delta R,t) \Omega(R+\Delta R).
\end{aligned}
\end{equation}
Taking the limit $\Delta R \to 0$, the angular momentum conservation equation becomes:
\begin{equation}
\label{eq:angular_cons_1}
    \frac{\partial}{\partial t} (\Sigma R^2 \Omega) + \frac{1}{R} \frac{\partial}{\partial R} \left( \Sigma R^3 \Omega v_R \right) = -\frac{3}{2R} \frac{\partial}{\partial R} \left( \nu \Sigma R^2 \Omega \right).
\end{equation}
Combining Eqs.~\ref{eq:mass_cons} and \ref{eq:angular_cons_1}, we recover the classical surface density evolution equation for a viscously spreading ring:
\begin{equation}
    \frac{\partial \Sigma}{\partial t} = \frac{3}{R} \frac{\partial}{\partial R} \left( \sqrt{R} \frac{\partial}{\partial R} \left( \nu \Sigma \sqrt{R} \right) \right).
\end{equation}

The EY effect introduces an additional torque term $ T_\EY $ on the right-hand side of Eq.~\ref{eq:angular_cons}. 
We begin by computing the magnitude of this torque.
 
The EY force, {defined as the thermal force component along the planetocentric orbital velocity vector averaged over a planetocentric orbit}, acting on a single ring particle at {an arbitrary} layer with the normal optical depth $\tau'$ (note that $\tau'$ is different from $\tau$ that accounts for the full thickness) is given by \citet{Zhou2024a}:
\begin{equation}
\label{eq:F_EY}
    F_\EY = \frac{(1 - A_{\rm v}) \,  \Phi_{\rm s} \pi r^2}{c} \, e^{-\tau'/\sin \psi} \, f_\EY,
\end{equation}
where $ \Phi_{\rm s} $ is the stellar flux (e.g., for the Sun, $ \Phi_{\rm s} \approx 1365~\rm W\,m^{-2} $ at 1~au), $A_{\rm v}$ is the particle's albedo in the visible band, $ r $ is the particle radius, $ c $ is the speed of light and $f_\EY$ is the dimensionless Eclipse–Yarkovsky coefficient. The angle $ \psi $ is the inclination of the ring plane with respect to the heliocentric orbital plane. If the ring has a dynamical optical depth $\tau$ and the Sun stands at elevation $\psi$ above the ring plane, the ray path through the particle layer is lengthened by $1/\sin \psi$, giving a suppression factor $ e^{-\tau/\sin \psi} $ of the radiation flux. In fact, we have $\sin \psi = \sin \varepsilon_{\rm p} |\cos \lambda|$ where $\varepsilon_{\rm p}$ is the planet's obliquity (e.g. $\varepsilon_{\rm p} \simeq 26.7^\circ$ for Saturn) and $\lambda$ is the orbital longitude \actaken{(starting with $\lambda=0$ at a solstice)}{}. The planet's obliquity could evolve due to planetary perturbation \citep[e.g. ][]{Ward1973, Ward2004, Hamilton2004, Saillenfest2021, Cuk2024}, which is ignored in this work for simplicity. The ring is assumed to be located in the planet's equatorial plane. 

In general, $f_{\EY}$ is a complex function of the rotation state, represented by the spin frequency $\omega$ and the obliquity $\varepsilon$ (i.e. the angle between the \actaken{particle}{} spin vector and the \actaken{pole of the orbit of the particle around the planet}{orbit pole}), the thermal properties, planet radius $r_{\rm p}$, particle radius $r$ and stellar irradiance. The full solution of EY coefficient obtained from solving the 3-D heat conduction equation for a spherical particle can be subdivided into the planetary diurnal component $f_{\EY,{\rm d}}$ (caused by the spin) and the planetary seasonal component $f_{\EY,{\rm s}}$ (caused by the {planetocentric orbit}) \citep{Zhou2024a}:
\begin{equation}
\label{eq:f_EY}
    f_\EY = f_{\EY,{\rm d}} + f_{\EY,{\rm s}} ,
\end{equation}
where 
\begin{align}
    f_{\EY,{\rm d}} &= \phantom{-}{4  \over 9  } \eta_{\rm shadow} \left[V(z_{\gamma-1}) \cos^4{\varepsilon \over 2} - V(z_{\gamma + 1})\sin^4{\varepsilon \over 2}\right], \label{eq:f_EY_d}\\
    f_{\EY,{\rm s}}&= - {2  \over 9} \eta_{\rm shadow}  V(z_1)\,\sin^2{\varepsilon} \label{eq:f_EY_s}.
\end{align}
Here \actaken{$\gamma = \omega/\Omega$, $z_x = \sqrt{-{\imath }x}\,r / r_\Omega$,}{$z_{ \gamma \pm 1 } = \sqrt{-{\imath }(\gamma \pm1)}\,r / r_\Omega$  with $\gamma = \omega/\Omega$, $z_1 = \sqrt{-{\imath }}\,r / r_\Omega $} and $\imath \coloneqq \sqrt{-1}$. The time fraction in the shadow $\eta_{\rm shadow}$ averaged over the planetocentric and heliocentric orbits can be approximated as
\begin{equation}
\label{eq:eta_shadow}
    \eta_{\rm shadow} = 0.5\, \left( {R  \over r_{\rm p}} \right)^{-2.1},
\end{equation}
for $\varepsilon_{\rm p} = 26.7^\circ$ in Saturn's case, which is validated in Appendix~\ref{app:eta_shadow}.
The thermal penetration depth over an orbit is $r_\Omega = \sqrt{K/(\rho C \Omega) }$ where $K$ and $C$ are the thermal conductivity and the specific heat capacity. We set the thermal conductivity $K = 10^{-4} ~\rm W\,m^{-2}\, K^{-1}$ and the specific heat capacity $C = 820~\rm J\,K^{-1} \, kg^{-1}$ \citep{Vokrouhlicky2007}. \actaken{Finally}{Here} $V(z)$ is a real-value function defined by 
\begin{equation}
    V(z) =  \operatorname{Im}{\left( 1 + \chi\, {z \over j_1(z)} {{\rm d}j_1(z) \over {\rm d}z}\right)^{-1}}
\end{equation}
with $j_1(z)$ denoting the spherical Bessel function of the first kind and order $1$
\begin{equation}
    j_1(z) = {\sin z \over z^2} - {\cos z \over z}.
\end{equation}
The variable $\chi$ is defined as
\begin{equation}
    \chi = {K \over \sqrt{2} r \epsilon \sigma T_{\rm sub}^3 (1 - r_{\rm p}/\pi R)^{3/4}}.
\end{equation}
Here $\epsilon$ is the emissivity of particle, $\sigma = 5.67 \times 10^{-8} \rm \,W~m^{-2}~K^{-4}$ is the Stefan-Boltzmann constant, $\eta$ is the thermal inertia, and $T_{\rm sub} = ((1-A_{\rm v})\Phi_{\rm s}/\epsilon \sigma)^{1/4} $ is the subsolar temperature with $A$ being the albedo.


For the sake of easier integration, we approximate the dependence of $f_\EY$ on the particle radius $r$ as:
\begin{equation}
\label{eq:f_EY_size}
f_\EY =
\begin{cases}
f_{\EY, 0}, & r \ge r_0, \\
f_{\EY, 0} \,\left( {r/r_0} \right)^3, & r < r_0,
\end{cases}
\end{equation}
where $f_{\EY,0}$ is only independent of  $r$, but still depends on other parameters such as rotational and thermal properties. Here $ r_0 \sim r_{\Omega}\sim \rm 1~mm $ is the characteristic thermal penetration depth, below which the thermal wave penetrates the body so that the surface temperature tends to be more uniform for smaller particles. A detailed justification for this approximation is provided in Appendix~\ref{app:f_EY}. Equation~\ref{eq:f_EY_size} is based on the assumption that particles of all sizes have the same rotation state. However, when we consider a ring system, we need to account for the size-dependent rotation distribution. We will show in Sec.~\ref{sec:numerical} that, even under a realistic rotation distribution in different size ranges, the approximation in Eq.~\ref{eq:f_EY_size} remains valid for estimating the total EY torque on the ring. 

For a layer of ring annulus where the optical depth is \actaken{between $\tau'$ and $\tau'+\Delta\tau'$}{ $\tau'$} at radius $R$, the total number of particles in this layer is $N \Delta \tau'/\tau$ where $\tau = \sum_i \pi r_i^2 \Sigma/m_i$ is the dynamical optical depth and $ N = 2R \Delta R\, \tau/r^2$ is the total number of particles in the annulus. The total torque exerted by the EY effect on all particles at the layer is given by
\begin{equation}
\label{eq:T_EY1}
    T_\EY = \sum_i F_{\EY,i} R = \frac{(1 - A_{\rm v}) \,  \Phi_{\rm s} R}{c} \, e^{-\tau'/\sin \psi} \sum_i \pi r_i^2 f_{\EY,i},
\end{equation}
where the sum is taken over all particles in the annulus. We begin by assuming that all ring particles have the same radius $ r $, in which case the torque simplifies to
\begin{equation}
\begin{aligned}
\label{eq:T_EY2}
    T_\EY  &= \frac{(1 - A_{\rm v}) \,  \Phi_{\rm s} R}{c} \, e^{-\tau'/\sin \psi} \,{N \Delta \tau' \over \tau} \pi r^2 \bar{f}_\EY \\
    &= \frac{(1 - A_{\rm v}) \,  \Phi_{\rm s} R}{c} \, e^{-\tau'/\sin \psi} \, {2\pi R\Delta R } \Delta \tau' \bar{f}_\EY
\end{aligned}
\end{equation}
The quantity $ \bar{f}_\EY \coloneqq \sum_i f_{\EY,i} / N $ is the mean Eclipse–Yarkovsky coefficient for particles with the radius of $r$, which accounts for variation in spin states across the particle population.  The total EY torque is the integral over $\tau'$ from 0 to $\tau$:
\begin{equation}
\label{eq:T_EY_layer}
\begin{aligned}
    T_\EY  &= \int_0^{\tau} 2\pi R^2 \Delta R \, \frac{(1 - A_{\rm v}) \,  \Phi_{\rm s} \bar{f}_\EY}{c} \, e^{-\tau'/\sin \psi} {\rm d}\tau' \\
    &= 2\pi R^2 \Delta R \frac{(1 - A_{\rm v}) \,  \Phi_{\rm s} \bar{f}_\EY }{c}   (1-e^{-\tau/\sin \psi}) \sin \psi.
\end{aligned}
\end{equation}
For an extremely tenuous ring with $\tau \ll 1$, $T_{\rm EY} \sim 2 \pi R^2 \Delta R (1 - A_{\rm v}) \,  \Phi_{\rm s} {\bar f_{\rm EY}} \tau /c$, indicating the total torque is proportional to the total area of ring particles in the annulus $2 \pi R \Delta R \tau$. On the other hand, for a dense ring with $\tau \gg 1$, $T_{\rm EY} \sim 2 \pi R^2 \Delta R (1 - A_{\rm v}) \,  \Phi_{\rm s} {\bar f_{\rm EY}} \sin \psi /c$, indicating the total torque is proportional to the projected area of ring $ 2\pi R \Delta R \sin \psi$. Interestingly, when $r>r_0\simeq \rm 1~mm$, the total torque does not depend on particle size since the mean EY torque efficiency $\bar{f_{\EY}} = \bar{f_{\rm EY,0}}$ that is independent on the size according to Eq.~\ref{eq:f_EY_size}. 

In reality, ring particles follow a
 certain SFD, often approximated with a power-law \citep[e.g.,][]{Brilliantov2015}. To account for this, we replace $ \bar{f}_\EY $ with a size-free variable $ \bar f_{\EY,0} $, and encapsulate the size effects into a dimensionless size correction factor $ \eta_{\rm size} $. The torque then becomes
\begin{equation}
\label{eq:T_EY3}
    T_\EY = 2\pi R^2 \Delta R \, \frac{(1 - A_{\rm v}) \,  \Phi_{\rm s} \bar f_{\EY,0} \eta_{\rm size}}{c} \, (1-e^{-\tau/\sin \psi}) \sin \psi,
\end{equation}
where the detailed expression for $ \eta_{\rm size} $ is provided in Appendix~\ref{app:eta_size}. In fact, when the considered minimum size $r_{\rm min} > r_0 \sim 1~$mm,  we have $\eta_{\rm size} = 1$ regardless of the SFD. When $r_{\rm min} < r_0$, we have $\eta_{\rm size }< 1$. {A typical power-law differential size–frequency distribution follows ${\rm d}N \propto r^{-\alpha}$ with $-\alpha$ being the power index. As shown in Appendix~\ref{app:eta_size}, $\eta_{\rm size}$ is dominated by the largest particles for shallow size–frequency distributions ($\alpha < 3$) and by the smallest particles for steep distributions ($\alpha > 3$). At the critical slope $\alpha = 3$, contributions from both large and small particles are significant. Cassini observations suggest that $\alpha$ in Saturn’s main rings spans the range $2.7 \lesssim \alpha \lesssim 3.2$ \citep[see review in][]{Miller2024}. In this study, we adopt $\alpha = 3$, which is also supported by theoretical expectations \citep{Brilliantov2015}.}

{Since $\psi$ vary with time over the solar seasons,} the factor $(1 - e^{-\tau/\sin \psi}) \sin \psi$ should be averaged over $\lambda$ to calculate the secular effect of the EY torque. {While the average of $\sin \psi$ is exactly $\frac{2}{\pi}\sin \varepsilon_{\rm p}$, the one of the full expression} does not lead to a closed-form expression in elementary functions. {We offer the following approximation, correct to within less than $2\%$:}
\begin{equation}
\begin{aligned}
    g(\tau, \varepsilon_{\rm p}) &=  <(1 - e^{-\tau/\sin \psi}) \sin \psi> \\
    &\approx (1 - e^{-\tau/(2\sin \varepsilon_{\rm p} /\pi)}) {2 \sin \varepsilon_{\rm p} \over \pi}\cdot G(\tau, \varepsilon_{\rm p}),    
\end{aligned}
\end{equation}
{where $G$ is a fitting function of order unity:}
\begin{equation}
    G(\tau, \varepsilon_{\rm p}) = 1 - 2u e^{-u} / 7
\end{equation}
with
\begin{equation}
    u = \tau\times\left(\frac{100}{\varepsilon_{\rm p}[{\rm deg}]}+\frac{\varepsilon_{\rm p}[{\rm deg}]}{150}\right).
\end{equation}
{Note that for all $\tau$ and $\varepsilon_{\rm p}$, $0.89<G(\tau,\varepsilon_{\rm p})\leqslant 1$; in other words, discarding $G$ degrades the fit by $\sim10\%$.}


Including the EY torque $T_\EY$ in the angular momentum conservation equation (Eq.~\ref{eq:angular_cons_1}) and taking the limit $ \Delta R \to 0 $, we arrive at

\begin{equation}
\label{eq:angular_cons_2}
\begin{aligned}
    \frac{\partial}{\partial t}(\Sigma R^2 \Omega) + \frac{1}{R} \frac{\partial}{\partial R}(\Sigma R^3 \Omega v_R) &= -\frac{3}{2R} \frac{\partial}{\partial R}(\nu \Sigma R^2 \Omega) \\
    &\quad + \frac{R (1 - A_{\rm v}) \,  \Phi_{\rm s}}{c} \, \bar f_{\EY,0} \eta_{\rm size} \, {g(\tau, \varepsilon_{\rm p})}.
\end{aligned}
\end{equation}
Combining Eqs.~\ref{eq:mass_cons}, \ref{eq:angular_cons_2} and two factors $\eta_{\tau}$ and $\eta_{\rm p}$ that will be introduced in Sec.~\ref{sec:consideration_tau} and \ref{sec:consideration_pY}, we obtain the modified surface density evolution equation including the EY effect:
\begin{equation}
\label{eq:full_evolution}
\begin{aligned}
\frac{\partial \Sigma}{\partial t} & = \frac{3}{R} \frac{\partial}{\partial R} \left( \sqrt{R} \frac{\partial}{\partial R} (\nu \Sigma \sqrt{R}) \right) \\
&- \frac{2 (1 - A_{\rm v}) \,  \Phi_{\rm s}  \eta_{\rm size}}{c R} \, \frac{\partial}{\partial R} \left( \frac{ \bar f_{\EY,0} R \, \eta_{\tau} \, \eta_{\rm p} \, {g(\tau , \varepsilon_{\rm p})} }{\Omega} \right).
\end{aligned}
\end{equation}
Here $f_{\rm EY,0}$ scales as $R^{-2.1}$ for $\varepsilon_{\rm p} = 26.7^\circ$ (Eq.~\ref{eq:eta_shadow}). Note that even in the case $\eta_{\rm size} = 1$, the particle size $r$ is still relevant in Eq.~\ref{eq:full_evolution}, as $r$ emerges from $\tau \sim 3 \Sigma/4\rho r$ in the EY term. Generally, the EY effect is inversely proportional to the particle size. The value of $f_{\rm EY,0}$ will be determined in Sec.~\ref{sec:numerical}.

\subsection{Consideration of high optical depth}
\label{sec:consideration_tau}

The analytic expressions for the EY effect (Eqs.~\ref{eq:f_EY}, \ref{eq:f_EY_d}, and \ref{eq:f_EY_s}) are derived under the assumption that each ring particle maintains a constant shape and spin state over the course of an orbital period. This assumption may be violated in dense regions of the ring, where high optical depths allow for frequent inter-particle collisions and the formation of transient non-spherical aggregates. To account for this, we introduce a conservative attenuation scheme to justify the implementation of the EY torque under such conditions.

In regions of high optical depth, ring particles can temporarily self-aggregate due to self-gravity and cohesion \citep[e.g.][]{Perrine2012}, forming gravitational wakes which are unstable, anisotropic structures that persist for roughly an orbital timescale. These wakes violate the assumption of spherical, isolated particles and significantly alter both the effective shape and emission geometry, which are critical for the EY effect. Since the analytic EY solution assumes a rigid, spherical body, its application becomes questionable in the presence of gravitational wakes. A common stability criterion for wake formation is the Toomre parameter $Q>2$ where
\begin{equation}
    Q = \frac{\Omega v_R}{3.36\, G \Sigma},
\end{equation}
where $ v_R $ is the radial velocity dispersion. Substituting $ v_R \sim \sqrt{ G m / r} $ and using the standard definition of optical depth $ \tau = 3\Sigma / (4 \rho r) $, we obtain:
\begin{equation}
    \tau < \frac{\rho_{\rm p}}{\rho}  \frac{\sqrt{4\pi G \rho /3}}{\Omega},
\end{equation}
where $ \rho_{\rm p} $ is the bulk density of the planet and $ \rho $ is the bulk density of ring particles. As an illustrative example, let us consider Saturn’s A ring. Taking Saturn’s density $ \rho_{\rm p} = 687\, \rm kg\,m^{-3} $, particle density $ \rho = 1000\, \rm kg\,m^{-3} $, and orbital period $ 2\pi/\Omega \approx 10 $ hours, we find $ \tau < 2 $, which is representative of typical planetary ring conditions.

In addition, rotation may be changed by collisions before the particle completes an orbital period in the dense ring. Since the EY torque depends sensitively on a particle’s rotation, the analytic expressions are only valid if the rotation state remains approximately constant during an orbit. While a fully time-variable spin might still produce a net EY torque, such behavior is beyond the scope of this work. To constrain this regime, we estimate the average number of collisions per particle per orbit as $4\pi \tau$. Since each collision would change the rotation of two particles, requiring fewer than one spin change per orbit per particle yields a critical optical depth of $ \sim 1/8\pi \simeq 0.05$.

Although a full treatment of EY dynamics under frequent collisions and wake-dominated conditions is deferred to future work, we conservatively model the suppression of the EY effect at high $\tau$ by multiplying a smooth attenuation function:
\begin{equation}
\label{eq:eta_tau}
\eta_{\tau} =
\begin{cases}
1, & \tau \le \tau_1 \\
(\tau_2 - \tau)/(\tau_2 - \tau_1 ), &\tau_1 < \tau < \tau_2 \\ 
0 &  \tau \ge \tau_2
\end{cases}
\end{equation}
This function captures the gradual decay of EY effectiveness with increasing optical depth and serves as a practical approximation in the absence of a full numerical model. In the illustrating examples in this paper, we adopt $\tau_1 = 1/8\pi \sim 0.05$ and $\tau_2 = 2$ for estimate. 

Note that the suppression of the EY effect at high $\tau$ does not imply that the EY effect is dynamically irrelevant in dense rings. In fact, the EY effect remains active at the ring edges where the optical depth is low, so that it can influence the overall angular momentum transport and thereby the global evolution of dense rings, as demonstrated in Sec.~\ref{sec:evolution}.

\begin{figure*}
    \centering
    \includegraphics[width=\linewidth]{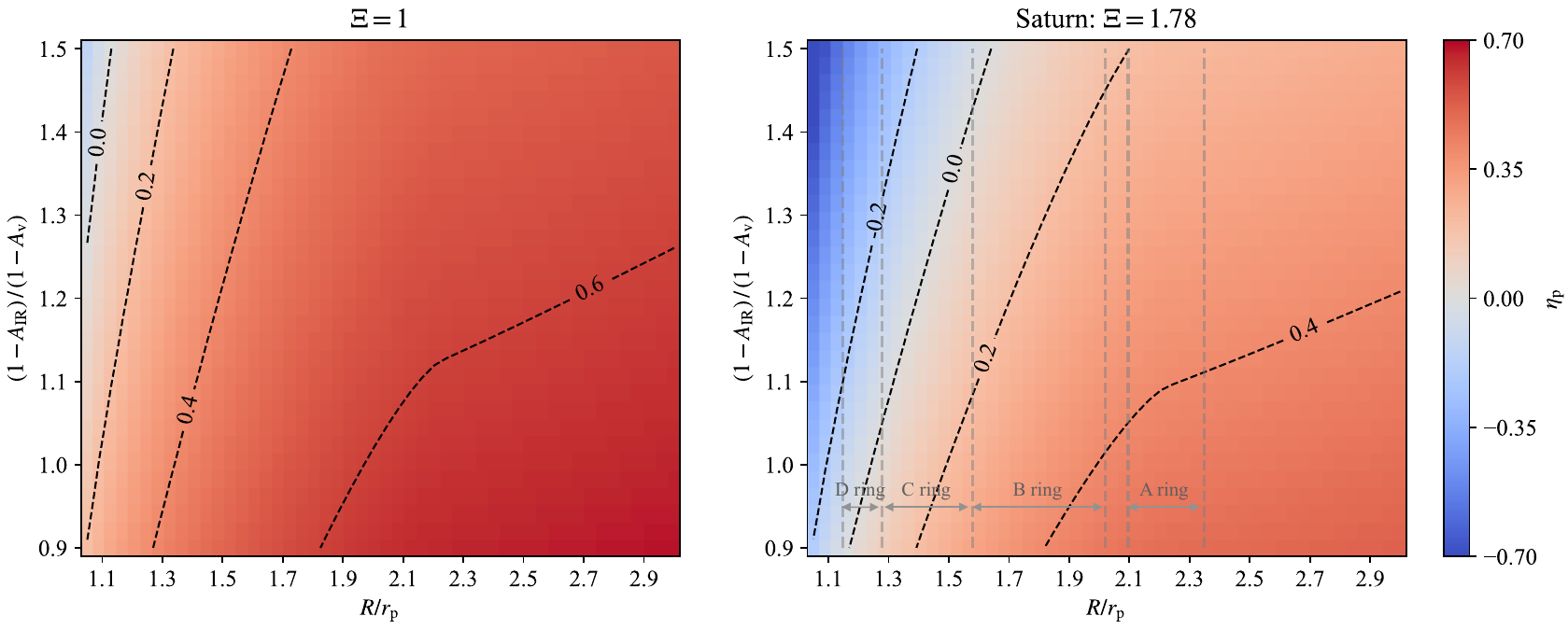}
    \caption{{The factor $\eta_{\rm p}$ due to the planetary radiation as a function of the orbital distance $R$ and $(1 - A_{\rm IR})/(1-A_{\rm v})$ (a higher value represents a higher influence of planetary radiation), considering different planet's emissivity $\Xi = 1$ (e.g. Martian case) and 1.78 (Saturn's case). The emissivity can partially come from the internal energy of the planet. For comparison among results obtained with different values of $\Xi$, we fix the planetary obliquity at $\varepsilon_{\rm p} = 25^\circ$, representative of both Mars ($25.2^\circ$) and Saturn ($26.7^\circ$). The planet's albedo is set as $A_{\rm p} = 0.342$. For Saturn's rings, previous works adopt $(1 - A_{\rm IR})/(1-A_{\rm v}) = 1.33$ for A/B rings \citep{rub2006, Vokrouhlicky2007}, while $A_{\rm IR}$ is highly uncertain.}}
    \label{fig:eta_p}
\end{figure*}

\subsection{Consideration of planetary radiation}
\label{sec:consideration_pY}

{Planetary radiation on the ring particle can produce an opposite effect to the EY effect \citep{rub2006, Vokrouhlicky2007, Zhou2024a}, which is termed as ``Yarkovsky-Rubincam effect'' \citep{Farinella1996}, ``Rubincam effect'' \citep{Cuk2005}, or ``planetary-Yarkovsky effect'' \citep{Vokrouhlicky2007}. 
\actaken{It has been shown \citep{rub2006, Vokrouhlicky2007} that the ratio between the thermal force caused by \edit1{the planetary radiation and that from the EY effect} can be written as
\begin{equation}
    \mathcal{F} =  {\bar \Phi_{\rm p} \over  (1 - A_{\rm v}) \,  \Phi_{\rm s}  \, \eta_{\rm shadow}(R, \varepsilon_{\rm p})\, g(\tau, \varepsilon_{\rm p})}.
\end{equation}
where $\bar \Phi_{\rm p}$ is the planetary radiation flux on the ring particle.}{} Interestingly, $\mathcal{F}$ is independent of particle properties, depending solely on the geometry of the ring.
This ratio yields a reduction factor $\eta_{\rm p}$\actaken{$=1-\mathcal{F}$}{} to $f_{\rm EY}$ (Eq.~\ref{eq:full_evolution}), with $\eta_{\rm p} = 1$ representing zero planetary radiation. Note that $\eta_{\rm p}$ could be negative and reverse the total torque if the planetary radiation is sufficiently strong. The effect of planetary radiation can be strengthened if (1) the ring particle's albedo in the wavelength range of the planet's thermal emission is low, which means the particle absorbs more planet's radiation; or (2) the planet emits more radiation than it receives due to the release of internal energy. \actaken{}{Here, we discuss the relative strength of the effect of planetary radiation to the EY effect.}}

\actaken{}{{It has been shown \citep{rub2006, Vokrouhlicky2007} that the ratio between the thermal force caused by the EY effect and that from the planetary radiation can be written as
\begin{equation}
    \mathcal{F} =  {\bar \Phi_{\rm p} \over  (1 - A_{\rm v}) \,  \Phi_{\rm s}  \, \eta_{\rm shadow}(R, \varepsilon_{\rm p})\, g(\tau, \varepsilon_{\rm p})}.
\end{equation}
Here $\bar \Phi_{\rm p}$ is the planetary radiation flux on the ring particle.  Therefore, the factor $\eta_{\rm p}$ attached to the EY torque is}
\begin{equation}
    \eta_{\rm p} = 1 - \mathcal{F}.
\end{equation}}

{\actaken{}{As the ring is separated from the planet within a few planetary radii, the planet cannot be treated as a point source, but needs to consider finite angular extent as seen from the ring.} In the following, we calculate \actaken{$\bar \Phi_{\rm p}$}{} the radiation received by a ring particle at the orbital distance $R$ with normal (dynamical) optical depth $\tau$\actaken{. As the ring is separated from the planet within a few planetary radii, the planet cannot be treated as a point source but needs to be decomposed in surface elements $\Delta S_p$.}{ from an arbitrary planetary surface element} \actaken{We assume the planet emits as a Lambertian surface of radiance $L$}{}.}

{The position vector of the ring particle is $\mathbf R$ and the position vector of the \actaken{}{an} arbitrary planetary surface element is $\mathbf r_{\rm p} = r_{\rm p}\,\hat{\mathbf n}_p$, where $\hat{\mathbf n}_p$ is both the outward surface normal and a unit vector on the sphere. The vector from the planet surface element to the ring point is
\begin{equation}
\mathbf s = \mathbf R - \mathbf r_{\rm p} ,
\end{equation}
and the unit vector is $\hat{\mathbf s} = \mathbf s / s $. A planetary surface element contributes only if it is visible from the ring point. This requires the Lambert emission cosine}
\begin{equation}
\mu_{\rm p} \equiv \hat{\mathbf n}_{\rm p} \cdot \hat{\mathbf s} > 0 .
\end{equation}
\actaken{}{

{We assume the absorbed stellar flux is redistributed uniformly over the planet and re-emitted thermally, and the planet emits as a Lambertian surface, characterized by a radiance} 
\begin{equation}
    L = {\Phi_{\rm s} \over  4\pi}
\end{equation}
{where the factor of $4$ comes from the uniform emission of the received radiation ($\pi r_{\rm p}^2 / 4\pi r_{\rm p}^2$).} }
{\actaken{Under this condition, the}{Consider a planetary surface element $\Delta S_p$ emitting radiance $L$. The} differential absorbed flux at a ring point, per unit ring midplane area, is
\begin{equation}
\label{eq:Delta_Phi_abs}
\Delta \Phi_{\rm abs}= L\, \mu_{\rm p}\, \frac{\Delta S_p}{s^2}\, \sin \psi_{\rm p}\, \left( 1 - e^{-\tau/\sin \psi_{\rm p}} \right),
\end{equation}
\actaken{Here, }{subject to the visibility conditions $\mu_{\rm p}>0$. Here $\mu_{\rm p}$ is the Lambert emission cosine of the planet patch, and} $\psi_{\rm p}$ is the elevation angle of radiation from the planetary surface element as seen from the ring plane (the same consideration as starlight elevation angle $\psi$ in Eq.~\ref{eq:T_EY_layer}), which is defined as,
\begin{equation}
\sin \psi_{\rm p} = \left| (-\hat{\mathbf s}) \cdot \hat{\mathbf n}_{\rm ring} \right| ,
\end{equation}
where $\hat{\mathbf n}_{\rm ring} $ is the normal vector of the ring plane, equal to the spin axis of planet.}

\actaken{{\actaken{Let us now estimate $L$. Assuming}{We assume} the absorbed stellar flux is redistributed uniformly over the planet and re-emitted thermally, \actaken{one gets for the thermal emission}{and the planet emits as a Lambertian surface, characterized by a radiance}} 
\begin{equation}
    L_{\rm IR} = {\Phi_{\rm s} (1-A_p) \over  4\pi}\Xi
\end{equation}
{where the factor of $1/4$ comes from the uniform emission of the received radiation ($\pi r_{\rm p}^2 / 4\pi r_{\rm p}^2$)}}{.} {, $1/\pi$ is the radiant intensity, power per unit solid angle, from  $A_p$ is the albedo of the planet in the visible, and $\Xi$ is a factor accounting for the fact that the planet may emit more than it receives from the sunlight due to the release of internal energy.}{}{Moreover, the planet reflects the visible light: $\displaystyle L_v={\Phi_s A_p \over 4}$ on average over the surface.}

{\actaken{}{In fact, not all of the planetary radiation can be absorbed by the particle. Assuming the planet's visible albedo is $A_{\rm p}$, the planet reflects $A_{\rm p}$ of the total solar flux in the visible band and absorbs $ (1 - A_{\rm p})$ and then re-emits in the infrared band (e.g. Saturn's radiation peaks at $\sim 30~\mu$m). In addition, the planet can emit more than it receives from the sunlight by a factor of $\Xi$ due to release of the internal energy.} Assuming the particle has a visible albedo $A_{\rm v}$ and an albedo $A_{\rm IR}$ in the planet's emission band, the total radiation absorbed by the particle is \actaken{$(1-A_{\rm v})L_v+(1-A_{\rm IR})L_{\rm IR}=\frac{\Phi_s}{4}[$}{ }$A_{\rm p} (1-A_{\rm v}) + \Xi (1-A_{\rm p}) (1 - A_{\rm IR})$\actaken{$]$}{}. Therefore, Eq.~\ref{eq:Delta_Phi_abs} \actaken{can be expanded}{should be corrected} as}
\begin{equation}
\begin{aligned}
\label{eq:Delta_Phi_abs_2}
\Delta \Phi_{\rm abs}= & {\Phi_{\rm s} \over 4 \pi}\, \mu_{\rm p}\, \frac{\Delta S_{\rm p}}{s^2}\, \sin \psi_{\rm p}\, \left( 1 - e^{-\tau/\sin \psi_{\rm p}} \right) \\
& \cdot \left[A_{\rm p} (1-A_{\rm v}) + \Xi \,(1-A_{\rm p}) (1 - A_{\rm IR}) \right].
\end{aligned}
\end{equation}

{\actaken{The total planetary flux absorbed by the ring particle is then $\bar \Phi_{\rm p} \ = \actaken{\iint}{\sum} \Delta \Phi_{\rm abs}$.}{ The surface integral is most conveniently evaluated numerically using Monte Carlo uniform sampling on the planet's surface.} The surface element area is $\Delta S_{\rm p} = r_{\rm p}^2 \sin \theta_{\rm p} \Delta \theta_{\rm p} \Delta \phi_{\rm p}  $ with $\theta_{\rm p}$ and $\phi_{\rm p}$ \actaken{}{are} the colatitude and azimuthal angles, respectively. \actaken{The surface integral is most conveniently evaluated numerically using Monte Carlo uniform sampling on the planet's surface.}{The total planetary flux absorbed by the ring particle is therefore}}
\actaken{}{\begin{equation}
    \bar \Phi_{\rm p} \ = \actaken{\iint}{\sum} \Delta \Phi_{\rm abs}.
\end{equation}}
\actaken{Noting \begin{equation}
    h(R, \tau) = \actaken{\iint}{\sum} \, \, \frac{ \mu_{\rm p}}{4 \pi s^2}\, \sin \psi_{\rm p}\, \left( 1 - e^{-\tau/\sin \psi_{\rm p}} \right) \Delta S_p,
\end{equation}}{}

{\actaken{the}{The} ratio $\mathcal{F}$ becomes
\begin{equation}
    \mathcal{F} = { h(R, \tau) \over \eta_{\rm shadow}(R, \varepsilon_{\rm p}) g(\tau, \varepsilon_{\rm p})} {A_{\rm p} (1-A_{\rm v}) + \Xi \,(1-A_{\rm p}) (1 - A_{\rm IR}) \over 1 - A_{\rm v}}
\end{equation}
\actaken{}{with}} 
\actaken{}{\begin{equation}
    h(R, \tau) = \iint \, \, \frac{ \mu_{\rm p}}{4 \pi s^2}\, \sin \psi_{\rm p}\, \left( 1 - e^{-\tau/\sin \psi_{\rm p}} \right) \Delta S_p.
\end{equation} }

{Figure~\ref{fig:eta_p} shows the factor $\eta_{\rm p}$\actaken{$=1-\mathcal{F}$}{} as a function of the orbital distance $R$ and $(1 - A_{\rm IR})/(1-A_{\rm v})$, accounting for $\Xi = 1$ (Martian case) and $\Xi = 1.78$ (Saturn's case). The optical depth $\tau $ is set to be 0.1 and the obliquity is $25^\circ$. Generally, $\eta_{\rm p}$ decreases as $(1 -A_{\rm IR} )/ (1-A_{\rm v})$ increases, since this ratio represents the ratio of the ability to absorb the thermal radiation from the planet to the visible radiation from the sun. On the other hand, $\eta_{\rm p}$ increases as $R$ increases, since the planet gets closer to a point source whose radiation is heavily suppressed by self-shadowing such that only the inner edge is illuminated.} 

{For Saturn, the Bond albedo $A_{\rm p}$ is 0.342 and the emissivity $\Xi$ is 1.78, which results in an emission flux of 4.4~$\rm W~m^{-2}$ from Saturn's surface with the solar flux at Saturn $\Phi_{\rm s} = 15.0~\rm W~m^{-2}$. The visible albedo of Saturn's rings $A_{\rm v}$ varies among the major rings: 0.5-0.6 for A and B icy ring particles \citep{Morishima2010} and 0.2-0.3 for the C ring \citep{Porco2005, Spilker2018}. On the other hand, the thermal albedo at Saturn's emission peak ($\sim 30~\mu$m) is poorly constrained for Saturn's rings \citep[see review in ][]{Miller2024}. Previous studies adopt $A_{\rm IR} = 0.335$ which yields $(1-A_{\rm IR})/(1-A_{\rm v}) = 1.33$ for A and B rings \citep{rub2006, Vokrouhlicky2007}. In the Martian case, where $\Xi \simeq 1$, the ring predominantly migrates outward unless the ring particles possess an exceptionally high visible-to-thermal albedo ratio.}

{Another planetary source of radiation flux arises from the mutual heating of ring particles. In a ring with a random spatial distribution, this heating is approximately isotropic, which marginally elevates the mean particle temperature without significantly altering the surface temperature gradient. Consequently, under these assumptions, mutual heating does not introduce new thermal dynamics but simply shifts the mean temperature. While a more sophisticated model might move beyond the assumption of isotropy by accounting for specific particle coordinates, such analysis lies outside the scope of this study and is left for future investigation.}


\begin{figure*}
    \centering
    \includegraphics[width= \linewidth]{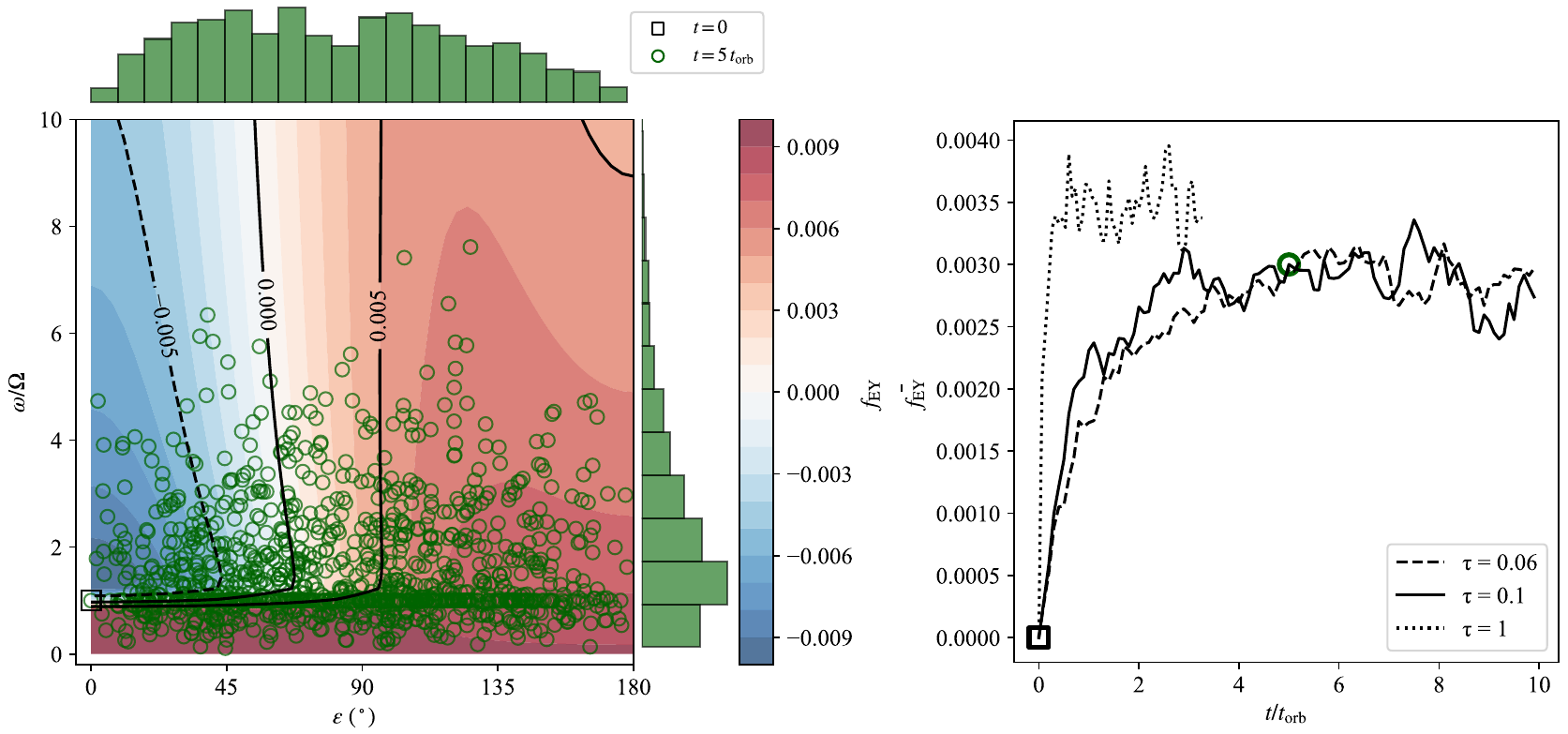}
    \caption{Illustration of the EY effect on the ring particles, taking the example of the Saturn ring. These ring particles are located at \grtaken{2}{1.5} Saturn radii, with the thermal inertia of $10$ tiu. The particle size ranges from 0.7~m to 2.3~m, distributed in a power law with the power index of -3. The left panel shows the rotation (i.e. the spin rate $\omega$ and the obliquity $\varepsilon$) distribution of ring particles at the $t = 0$ (black square) and $t = 5 ~t_{\rm orb}$ (green circles). The colors in the background denote the values of $f_{\rm EY}$. The right panel shows the convergence of the EY coefficient $f_{\rm EY}$ over time with the epochs in the left panel included, considering different optical depths $\tau$. This indicates that we can simply adopt a constant $\bar {f_{\rm EY}}$ over the long-term evolution.  }
    \label{fig:EY_spin}
\end{figure*}

\section{Numerical determination of the EY-coefficient}
\label{sec:numerical}

The direction and magnitude of the EY torque is represented by the mean dimensionless EY-coefficient $\bar f_{\EY, 0}$ (Eq.~\ref{eq:T_EY3}) of ring particles, which depends on the rotational states (i.e. $\omega$ and $\varepsilon$, Eqs.~\ref{eq:f_EY_d} and \ref{eq:f_EY_s}). 

The rotation distribution of particles is dominated by the rotational energy exchange between particles during mutual collisions \citep[e.g.][]{Salo1995, Morishima2004, Ohtsuki2005a, Ohtsuki2006a, Ohtsuki2006b}. Considering the rotational energy $E_{\rm rot} \propto r^{5} \omega^{2}$, the equipartition of rotational energy among particles tends to cause the spin rate to decrease with the size as $\omega \propto r^{-2.5}$. However, the spin rate is determined by the balance of two effects during collisions: (1) Collisional stirring, which transfers kinetic energy from random translational motion to rotational motion during collisions, tending to spin up smaller particles, with the spin rate scaling as $\omega \propto r^{-2.5}$; and (2) Rotational friction, which describes the dissipation of relative spin motions between colliding particles through frictional torques, tending to suppress spin differences and produce a spin rate that is nearly independent of particle size. The combination of these two effects leads to a spin rate scaling as $\omega \propto r^{-1}$ \citep{Ohtsuki2005a}. Given a size distribution following a power law ${\rm d}N \propto r^{-\alpha}$ with $\alpha = 3$, the mean spin rate follows 
\begin{equation}
\label{eq:omega_bar}
    \bar \omega \sim \Omega \left( \frac{r}{0.5 \, r_{\rm max}} \right)^{-1},
\end{equation}
with the maximum radius of the rings $r_{\rm max}$ being an arbitrary value \citep{Ohtsuki2005a}. On the other hand, the obliquities are random without any dependence on size.


To obtain the statistical distribution of $\omega$ and $\varepsilon$, we performed simulations of planetary rings using the \texttt{pkdgrav} gravitational $N$-body code \citep{Richardson00, Stadel01, Perrine2012}. Particle collisions were modeled using the soft-sphere discrete element method (SSDEM), in which particles can overlap as a proxy for surface deformation. The contact forces between overlapping particles are determined by the degree of overlap and the contact history using a spring dash-pot model, where reaction forces and torques are determined by a user-specified Hookes' law spring constant (related to the material stiffness), damping parameters (related to the coefficient of restitution) and coefficients of static, rolling, and twisting friction \citep{Schwartz12, Zhang17}. The SSDEM model allows for realistic modeling of low-speed collisions with finite contact durations, dissipation (via friction), and angular momentum transfer. The simulations are performed in a co-moving Hill frame for a small ``patch'' of ring particles. Self-gravity is included through the use of periodic boundary conditions with ``ghost'' particles that interact gravitationally with the patch. We refer the reader to \citet{Perrine2011} or \citet{Ballouz2017} for a detailed description of \texttt{pkdgrav}'s implementation of planetary rings.


The particle sizes are assumed to follow a differential power-law SFD with an index of $\alpha = 3$. The maximum-to-minimum ratio of the size is set to be $r_{\rm max}/r_{\rm min} = 3$ to reduce the computational cost of the nearest-neighbor search. Therefore, the expected mean spin rate is $\bar \omega \sim \Omega$ according to Eq.~\ref{eq:omega_bar}. 

Figure~\ref{fig:EY_spin} shows an example of the evolution of the rotational distribution of ring particles in the Saturn system, with the parameters listed in Appendix~\ref{app:sim_para}. The particles are assumed to have a mean radius of 1~m, located at $R = 2 \, r_{\rm p}$. The initial condition for every particle is $\omega = \Omega$ and $\varepsilon = 0 $, for which the EY effect disappears. It can be seen that after the time of $t = {\rm a ~few~} t_{\rm orb} < t_{\rm col} = t_{\rm orb} /\tau $, the rotational distribution (left panel) and the mean EY coefficient $\bar f_{\rm EY,0}$ (right panel) evolve into equilibrium states. The EY coefficient $f_{\rm EY,0}$ of each particle is calculated with the complete formula in Appendix~\ref{app:f_EY} and the assumed thermal properties. Typically, the spin rates normalized by the orbital frequency $\Omega$ follow a Maxwellian distribution peaked at 1, and the obliquities are distributed more evenly, consistent with the prediction of Eq.~\ref{eq:omega_bar}.  The material properties,such as the friction angle and restitution coefficient (Table~\ref{tab:sim_params}), are adopted from asteroid studies \citep{Zhang20, Zhang2022}. Changing material properties may also influence the resulting spin distribution, which could be interesting to explore in future work. The EY coefficient converges to $\sim 0.003$, as shown in Fig.~\ref{fig:EY_spin}. 

With this rotational distribution, we extrapolate the results down to particle sizes as small as 1~mm, according to Eq.~\ref{eq:omega_bar}. Below this threshold, the EY effect diminishes rapidly and contributes negligibly to the total torque (Eq.~\ref{eq:f_EY_size}). We found that the EY coefficient $f_\EY$ remains positive with the range from $3 \times 10^{-3}$ to $9\times 10^{-3}$. Therefore, we conclude that the total EY effect for ring particles following a realistic SFD is positive and we approximate it as $ \sim 3 \times 10^{-3}$ for calculation in this paper.

We also explore the rotational distribution and the resulting $\bar f_{\EY, 0}$ by varying the optical depth, as shown in the right panel of Fig.~\ref{fig:EY_spin}. We found that the rotational distribution remains the same, and $f_{\EY, 0} \sim \grtaken{0.003}{0.005}$ at $R = 2\,r_{\rm p}$ is independent of $\tau$. This slight difference at $\tau = 1$ may be caused by the gravitational wake structures due to self-gravity. As we discussed in Sec.~\ref{sec:formula}, the gravitational wake structure makes the presented EY estimate unreliable and would require new techniques to analyze. Therefore, combined with Eq.~\ref{eq:eta_shadow}, $\bar f_{\EY, 0}$ can be approximated as
\begin{equation}
    \bar f_{\EY, 0} = \grtaken{0.003}{0.005} \, \left( {R \over 2 \, r_{\rm p}} \right)^{-2.1},
\end{equation}
which can be used in Eq.~\ref{eq:full_evolution} for solving the ring evolution.

We conclude that the net EY torque on the ring is positive, accounting for rotation distribution of all ring particles under collisions. Note that other factors such as the thermal properties and orbital radius only affect the magnitude, instead of the sign of $f_{\EY, 0}$ (see e.g. Eqs.~\ref{eq:f_EY_d} and \ref{eq:f_EY_s}). The outcome of the positive torque is not unexpected: both the seasonal component (Eq.~\ref{eq:f_EY_s}) and the second term in the diurnal component (Eq.~\ref{eq:f_EY_d}) are inherently positive, whereas the first term in the diurnal component can be either positive or negative. Therefore, due to more positive components, their combined contribution would probably result in a net positive torque, which is also illustrated by larger red area (positive $f_{\rm EY}$) in Fig.~\ref{fig:EY_spin}.

\section{Long-term evolution with the EY effect}
\label{sec:evolution}

\begin{figure*}
    \centering
    \includegraphics[width= \linewidth]{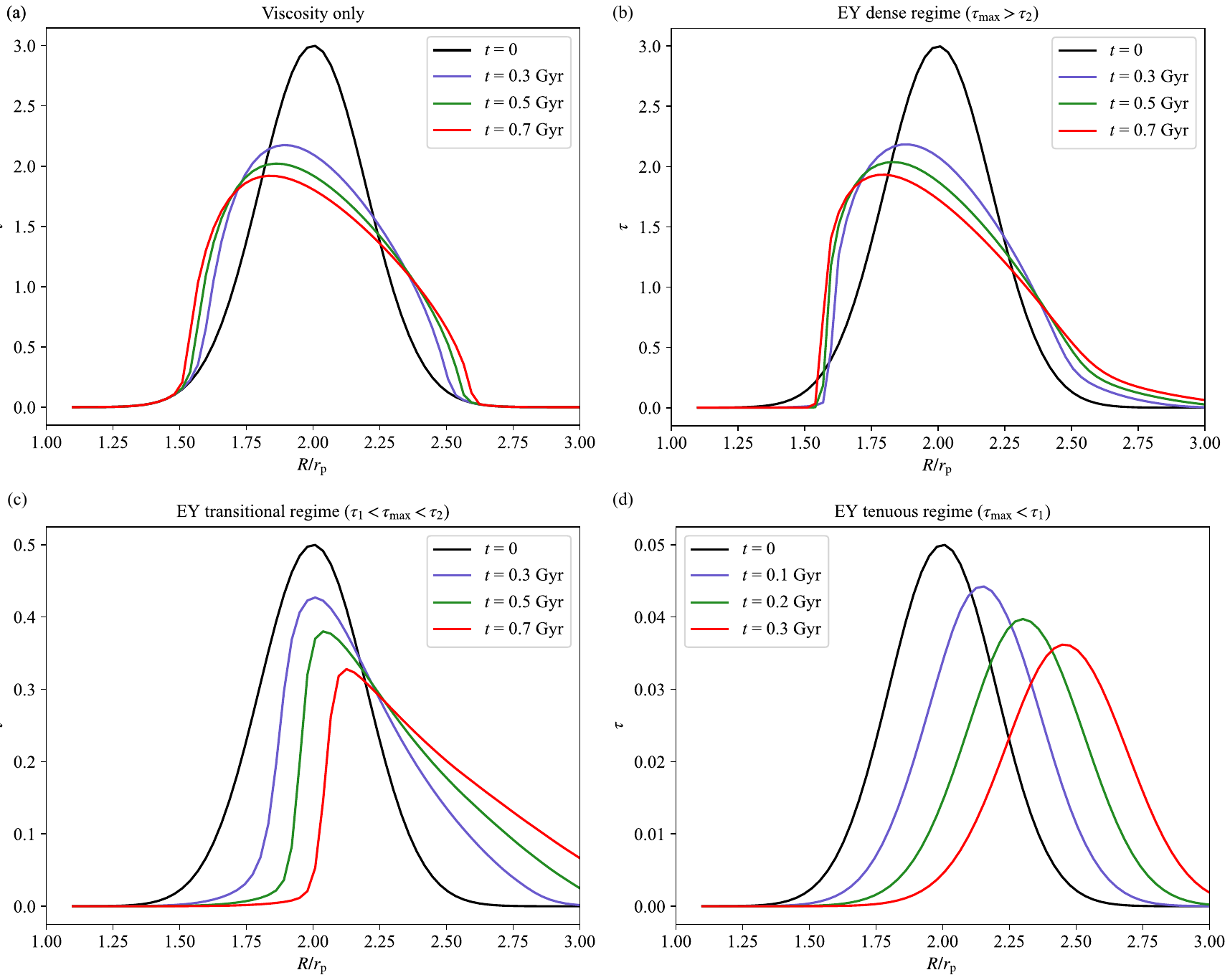}
    \caption{Ring evolutionary examples for the viscous evolution (a) and EY-included evolution in the order of decreasing the optical depth from (b) to (d). The simulation domain extends to $4\, r_{\rm p}$, while the panel displays to $3\, r_{\rm p}$, corresponding to the commonly adopted fluid Roche limit. The rings are assumed to orbit Saturn, with a mean particle size of 0.1~m. Note the difference in $y$-axis scales between the subfigures.}
    \label{fig:ring_evolution}
\end{figure*}

In this section, we investigate the long-term evolution of planetary rings with the viscous effect and the EY effect by numerically solving Eq.~\ref{eq:full_evolution} with a forward Euler method. We adopted zero surface density as the boundary conditions. As an illustration, we consider a synthetic Saturn-like system, that is, a planet with a mass of $5.7\times 10^{26}~$kg with the obliquity of $26.7\,^\circ$ with the ring lying in the equatorial plane. The particles are assumed to have an equal size of 0.1~m, with the thermal inertia of 10 tiu [$\mathrm{J\,m^{-2}\,K^{-1}\,s^{-1/2}}$]. {For the purpose of illustrating the EY effect, we adopt $\Xi = 1$, $A_{\rm p} = 0.3$ and $(1-A_{\rm IR})/(1-A_{\rm v}) = 1$ for the calculation of $\eta_{\rm p}$. Note that the exact case for Saturn (i.e. $\Xi = 1.78$, $A_{\rm p} = 0.342$ and $(1-A_{\rm IR})/(1-A_{\rm v}) = 1.33$ will be discussed in Sec.~\ref{sec:inner_edge}).} The viscosity can be approximated as \citep{Salmon2010}  
\begin{equation}
\nu =
\begin{cases}
 0.2 \, [{ {v_R^2 \tau /  \Omega (1 + \tau^2)} }] + r^2 \Omega \tau, & Q \ge 2 \\
 {1.4 \, (r_h^5 G^2 \tau^2 \rho^2 /  r^3 \Omega^3} )+ r^2 \Omega \tau, & Q < 2
\end{cases}
\end{equation}
with $r_{\rm H} = (2m/3m_{\rm p})^{1/3} R$ being the the mutual Hill radius of particles.

Figure~\ref{fig:ring_evolution} illustrates examples of ring evolution with viscosity dominated and EY dominated, obtained by varying their relative strengths and optical depth. The initial optical depth is assumed to follow a Gaussian profile. In the case where the EY torque can strongly affect the ring without considering any suppression (i.e. $\eta_\tau$ and $e^{-\tau/\sin \psi }$)\actaken{, the}{. The} evolutionary paths can be classified into three modes, based on $\eta_\tau$. These modes may be realized successively during the lifetime of a single ring as its surface density evolves.

Regime 1: Dense regime ($\tau_{\rm max} > \tau_2$). In regions where $\tau>\tau_2$, viscous transport dominates, driving material inward. However, this inward migration halts at the ring edge where $\tau < \tau_2$, as the EY torque becomes significant. At the edge, the positive EY torque pushes material outward until it is balanced by the opposing viscous torque that arises from particle accumulation. This interaction naturally produces a sharp edge. The overall evolution in this regime is thus characterized by an initial inward drift due to viscosity, followed by outward migration driven by the EY effect.

Regime 2: Transitional regime (${\tau_2}> \tau_{\rm max} > \tau_1$). In this intermediate regime, the EY effect is partially suppressed in regions with $\tau > \tau_1$, both by the factor $\eta_\tau$ (Eq.~\ref{eq:eta_tau}) and the exponential factor $e^{-\tau/\sin \psi}$ (Eq.~\ref{eq:F_EY}). Consequently, the inner regions of the ring—where the EY torque remains more efficient—migrate outward more rapidly than the denser outer portions, producing a pronounced sharp inner edge. Meanwhile, the outer edge stretches outward as the region closer to the outer edge has a lower optical depth and thus moves faster. The global evolution is dominated by net outward migration driven by the EY effect.

Regime 3: Tenuous regime ($\tau_{\rm max} < \tau_1$). When the entire ring is optically thin, the EY effect operates efficiently throughout the whole ring. The result is a nearly uniform outward expansion of the ring without major distortion of its surface density profile.

Note that these three regimes are classified based on the suppression factor $\eta_\tau$ and the assumption of a stronger EY effect than the viscous effect, given that the suppression factor $\eta_\tau$ is ignored. If, instead, the EY timescale is considerably longer than the viscous timescale, then the behavior characteristic of Regime 3 could also occur within a transitional ring or a tenuous ring; that is, the ring may first drift inward due to viscosity before reversing direction and migrating outward under the influence of the EY torque.

Overall, the principal outcome of the EY effect is to decrete the ring, driving material outward, rather than to accrete it inward, as expected under the classical viscosity-dominated picture. The degree to which a planet decretes its ring depends on the relative strength of the EY torque compared to viscous transport and on the local optical depth conditions described by the three regimes above.

\section{\actaken{Applications and discussion}{Discussion}}
\label{sec:discussion}

The EY effect introduces a positive angular momentum flux over long-term evolution, which has important implications for the origin, structure, and dynamical age of ring systems. This includes both planetary rings and asteroidal rings observed in the Solar System \citep[e.g.,][]{BragaRibas2014, Morgado2023}. We briefly discuss the applications to ring systems by examples of Saturn's rings and the Martian past ring.

\subsection{Inner sharp edges of Saturn's rings}
\label{sec:inner_edge}

\begin{figure}
    \centering
    \includegraphics[width=\linewidth]{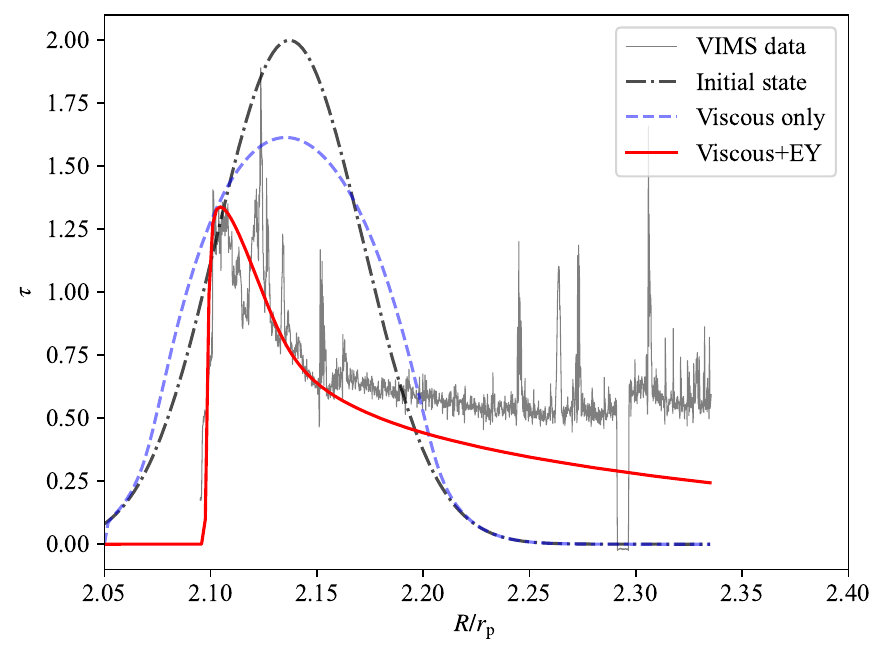}
    \caption{Saturn's A ring evolution with our model with only the viscous and EY effect included, in comparison with observation data. Here, the EY effect is estimated based on the particle size of 0.1~m and the particle thermal inertia of 10 tiu. The evolutionary time is 81 Myrs. The initial $\tau$ profile is a Gaussian profile. The simulation domain extends to $2.4\, r_{\rm p}$, while the panel displays to $\sim 2.33\, r_{\rm p}$, to be consistent with the outer edge of A ring confined by 7:6 inner Lindblad resonance with Janus/Epimetheus. }
    \label{fig:Saturn_A}
\end{figure}

Saturn’s A and B rings display remarkably sharp inner and outer edges which are not expected from viscous evolution. While the sharp outer edges are well explained by gravitational shepherding from nearby moons, the origin of the sharp inner edges remains an open question \citep[see, e.g.,][]{Crida2025}. One proposed solution is ballistic transport, in which high-speed micrometeoroid impacts redistribute mass inward and angular momentum outward from the inner edge, helping maintain a pre-existing sharp inner edge \citep{Estrada2023}. However, the sharp edge itself cannot be created by the ballistic transport, implying a missing mechanism that initially produces the sharp inner edge \citep{Estrada2023, Crida2025}.

As discussed in Sec.~\ref{sec:evolution}, both the EY transitional and dense regimes can spontaneously produce sharp inner edges. In particular, the shape of the edge of the bulk ring modelled in Figure~\ref{fig:ring_evolution} near $1.5 r_{\rm p}$ is reminiscent of the B-C ring transition.
In addition, our simple model demonstrates that the inner sharp edge and overall surface density profile of the A ring can be qualitatively reproduced in the EY-dominated dense regime, as shown in Fig.~\ref{fig:Saturn_A}. For this illustrative simulation, we adopt a particle size of 0.1~m with optical depth thresholds $\tau_1 = 0.05$ and $\tau_2 = 2$. {For the calculation of $\eta_{\rm p}$, we adopt $\Xi = 1.78$, $A_{\rm p} = 0.342$, and $(1 - A_{\rm IR})/(1 - A_{\rm v}) = 1.33$, following \citet{rub2006} and \citet{Vokrouhlicky2007}.} The initial ring profile is assumed to follow a Gaussian distribution. As this is a simplified model including only viscosity and the EY effect, it does not capture the finer structural details caused by gravitational interactions with Saturn or its moons, which are beyond the scope of this study.

\subsection{Long-term evolution of Saturn's rings}
\label{sec:C_D_ring}

Traditional viscous spreading transports the majority of mass inward, but the EY torque may produce an outward drift of Saturn's main rings {except for dense regions}. Saturn's C and D rings could be exceptions, as they are so close to Saturn that the EY torque could be even reversed by planetary radiation (Fig.~\ref{fig:eta_p}). Determination of the orbital evolution direction requires the precise value of the infrared albedo $A_{\rm IR}$ of particles in C and D rings. 

Saturn’s D ring is optically very thin $\tau < 10^{-3}$, and the particle sizes are believed to be 1-100 microns \citep{Hedman2007}. Hence, the Poynting--Robertson (P--R) drag may dominate \citep{Vokrouhlicky2007, Rubincam2013b, Hyodo2025} to drive materials inward, and the EY torque may be negligible.

In contrast, in the C ring, the particle sizes range from centimeters to meters \citep{Cuzzi2009, Hedman2011}, falling within the range where the EY effect is expected to be most efficient (i.e. larger than the thermal skin depth). The optical depth lies between 0.05 and 0.35, placing it within the tenuous or transitional regimes described in Sec.~\ref{sec:evolution}.
Saturn’s C ring is dynamically active and structurally complex, with structures like the ``plateaus” \citep[e.g.,][]{French2017} whose origin is poorly understood. \citet{Estrada2023} suggest that C ring structures could emerge from the effects of micrometeoroid-driven ballistic transport \citep{Durisen1989} onto the B ring. It may also interact with or counteract such an inward-acting process, potentially contributing to the formation. Therefore, the dynamical evolution and the resulting dynamical ages for Saturn's rings may require revisiting. These modifications to the standard dynamical model, including the interplay between EY torques and ballistic transport, will be explored in detail in a forthcoming companion study.



\subsection{Planetary moon formation}
\label{sec:satellite}

The EY effect introduces a novel pathway for satellite formation by enabling efficient outward angular momentum transport near ring edges. In classical models of viscous spreading, some ring material may diffuse across the Roche limit and gradually accumulate into moonlets \citep{Crida2012, Ciarniello2024, Blanc2025}. However, in low-density rings, viscous torques are often too weak to drive substantial mass outward before the ring is lost to planetary accretion. The EY torque, by contrast, can act more efficiently in tenuous regions and accelerate the outward drift of particles toward the Roche limit, especially at ring boundaries.

This mechanism offers a fresh perspective on how ancient rings may dissipate and contribute to satellite formation. For example, \citet{Hesselbrock2017} proposed a ring–moon cycling scenario for Mars, wherein rings repeatedly condense into moons that spiral inward and disrupt to form new rings. This model successfully reproduces the low mass and orbital characteristics of Phobos but predicts the existence of a co-existing tenuous ring, which has not been observed. A possible resolution is offered by the EY effect: the positive EY torque could expel the tenuous ring by pushing material beyond the Roche limit, facilitating its accretion onto Phobos.

\citet{Madeira2023} previously attempted to include the EY effect in this context but incorrectly modeled it as a negative torque. Our updated framework, incorporating the correct torque direction and magnitude, suggests that the EY effect may be a critical missing link in the completion of the ring–moon cycle. This scenario is the focus of a parallel study currently in preparation.

\section{Conclusions}
\label{sec:conclusion}

In this work, we have introduced and quantified the eclipse–Yarkovsky (EY) effect as a mechanism for the long-term dynamical evolution of planetary rings. This effect arises from the recoil force of ring particles due to asymmetric thermal emission following eclipse events, and its strength depends on particle size, thermal inertia, rotation state, and heliocentric distance. {The underlying eclipse-modulated Yarkovsky torque has previously been explored primarily for individual objects in binary or satellite systems, such as Earth satellites and binary asteroids, where it is commonly referred to as the Yarkovsky–Schach effect. It has also been discussed in the context of Saturn’s rings at the level of individual particles.}

{Here, we extend this effect to the collisionally-coupled planetary rings, which are considered as a continuous medium. Because inter-particle collisions efficiently transport angular momentum between neighbouring particles, the angular-momentum input at the particle scale can be treated as a flux acting on the ring annulus as a whole. This allows the EY effect to be incorporated self-consistently into a continuum description of ring evolution.} We derive the full surface-density evolution equation including both viscous and EY torques (Eq.~\ref{eq:full_evolution}).

{The planetary thermal radiation poses an opposite effect to the EY effect, whose strength depends on the planet’s albedo and internal energy release, as well as the ring particles’ albedos in the visible and thermal wavelength ranges. We quantify this contribution by numerically integrating the radiation force from each emitting surface element of the planet, assuming Lambertian emission.}

We validated the EY torque prescription using rotational distributions obtained from $N$-body simulations by code \texttt{pkdgrav}, showing that the EY effect produces a net positive angular momentum flux under typical conditions. Notably, we find that the total EY torque is independent of the particle size distribution and size range if the minimum size exceeds the thermal skin depth$\sim$1~mm (but decreases with a decreasing size below the thermal skin depth). Instead, the total torque depends primarily on the ring's optical depth.

Based on the optical depth, we identify three evolutionary regimes: tenuous, transitional, and dense regimes, depending on the optical depth. In the tenuous regime, the EY effect drives the whole ring outward with minor modification on the profile shape. In the transitional regime, the EY effect is gradually suppressed with the increasing optical depth, leading to a sharp inner edge due to the stronger EY effect in the tenuous edge region. The global drift is outward. In the dense regime, the EY effect can be overpowered in the densest region, resulting in the dominance of the viscous evolution. The sharp inner edge can also be produced in the balance of the EY effect and viscosity. The overall evolution in this regime is a viscous spreading, followed by a net outward migration driven by the EY effect.

Our numerical results show that the EY effect could effectively drive Saturn's rings outward, unless they are very close to Saturn (e.g. the location of the D ring), offering a new evolutionary pathway. Moreover, the EY effect spontaneously generates sharp inner edges, providing a natural explanation for features in Saturn’s rings previously attributed solely to micrometeoroid-driven ballistic transport. Importantly, the EY effect may also contribute to satellite formation by facilitating material transfer across the Roche limit, with application to Martian moon formation, which is under preparation in a parallel study. In conclusion, the EY effect is found to be a crucial mechanism affecting the formation, structure, and dynamical lifetimes of both planetary and asteroidal rings.

\bigskip
W.-H. Z. acknowledges support from the Japan Society for the Promotion of Science (JSPS) Fellowship (P25021). 
E.K.\ is supported by JSPS KAKENHI Grants No.\ 24K00698.
H.A. thanks CNES for support. 
The work of D.V. was supported by the Czech Science Foundation through grant 25-16507S. 
Numerical computations were carried out in part on the Licallo system at the Observatoire de la C\^ote d'Azur and on Cray XD2000 at the Center for Computational Astrophysics, National Astronomical Observatory of Japan.

\appendix



\section{The time fraction in shadow $\eta_{\rm shadow}$}
\label{app:eta_shadow}
We numerically calculated the time fraction in shadow for a particle at an orbital radius $R$. To determine whether a ring particle is in the planetary shadow, we compute the minimum distance between the planet's center and the line connecting the particle and the Sun. The planet is placed at the origin with radius $r_{\mathrm{p}}$, and the unit vector of the solar direction is defined in the equatorial plane as
\begin{equation}
    \hat{\mathbf{d}} = (\cos\phi,\, \sin\phi,\, 0),
\end{equation}
where $\phi$ is the heliocentric longitude of the Sun measured from the $x$-axis.
The position of a ring particle on a circular orbit of radius $R$ with an inclination $\varepsilon_{\rm p} = 26.73^\circ$ with respect to the heliocentric orbit is given by
\begin{equation}
    \hat{\mathbf{R}}(\xi) = (\cos\xi,\, \cos \varepsilon_{\rm p}\,\sin\xi,\, \sin \varepsilon_{\rm p}\,\sin\xi),
\end{equation}
where $\xi$ is the planetocentric true anomaly measured in the ring plane.  

A shadow occurs when this line intersects the planetary sphere of radius $R_{\mathrm{p}}$. 
This is equivalent to requiring that the minimum distance between the line and the origin is smaller than $R_{\mathrm{p}}$,
\begin{equation}
    \mathrm{dist}_{\min} = R\sqrt{\,1 - (\hat{\mathbf{R}}\cdot\hat{\mathbf{d}})^2\,} \le r_{\mathrm{p}},
\end{equation}
and that the intersection lies between the planet and the particle, i.e.\ along the direction from the particle toward the Sun,
\begin{equation}
    \hat{\mathbf{r}}\cdot\hat{\mathbf{D}} \le 0.
\end{equation}
The combined condition becomes
\begin{equation}
    \hat{\mathbf{R}}\cdot\hat{\mathbf{d}} \le -\sqrt{\,1 - \left( {r_{\rm p} \over R} \right)^2\,}.
\end{equation}
Substituting the explicit forms of $\hat{\mathbf{r}}$ and $\hat{\mathbf{D}}$, we obtain the condition:
\begin{equation}
\label{eq:shadow_judge}
    \cos\xi\,\cos\phi + \cos \varepsilon_{\rm p}\,\sin\xi\,\sin\phi
    \le -\sqrt{\,1 - \left( {r_{\rm p} \over R} \right)^2}.
\end{equation}

In the numerical algorithm, we set both $\xi$ and $\phi$ uniformly in the interval $[0, 2\pi]$.
For each pair $(\xi_i, \phi_j)$,  we check whether it satisfies the Ineq.~\ref{eq:shadow_judge}.
The total number of ``shadowed'' grid points, normalized by the total number of samples, gives the time-averaged shadow fraction:
\begin{equation}
    \eta_{\mathrm{shadow}}(R)
    = \frac{1}{(2\pi)^2}\iint H\!\left[-\cos\xi\,\cos\phi - \cos \varepsilon_{\rm p}\,\sin\xi\,\sin\phi - \sqrt{1 - \left( {r_{\rm p} \over R} \right)^2}\right]
    \, d\xi\, d\phi,
\end{equation}
where $H(x)$ is the Heaviside step function. This two-dimensional averaging procedure ensures that both the ring orbital motion and the planet’s orientation relative to the Sun are accounted for.

\begin{figure}
    \centering
    \includegraphics[width=\linewidth]{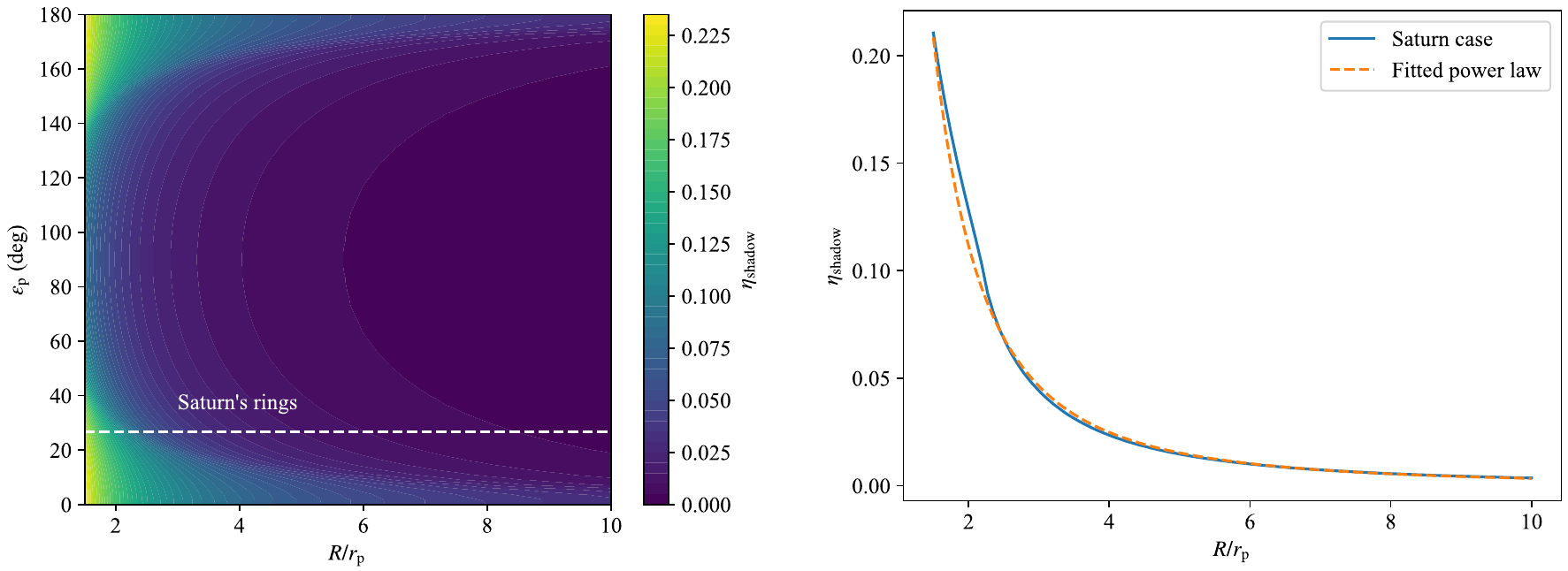}
    \caption{The time faction in shadow $\eta_{\rm shadow}$ as a function of the orbital radius $R$ in the Saturn's case where the planet obliquity is $\varepsilon_{\rm p} = 26.73^\circ$. The fitted function is $\eta_{\mathrm{shadow}} = 0.5 \,(R/r_{\mathrm{p}})^{-2.17}$. }
    \label{fig:eta_shadow}
\end{figure}

\begin{figure}
    \centering
    \includegraphics[width=\linewidth]{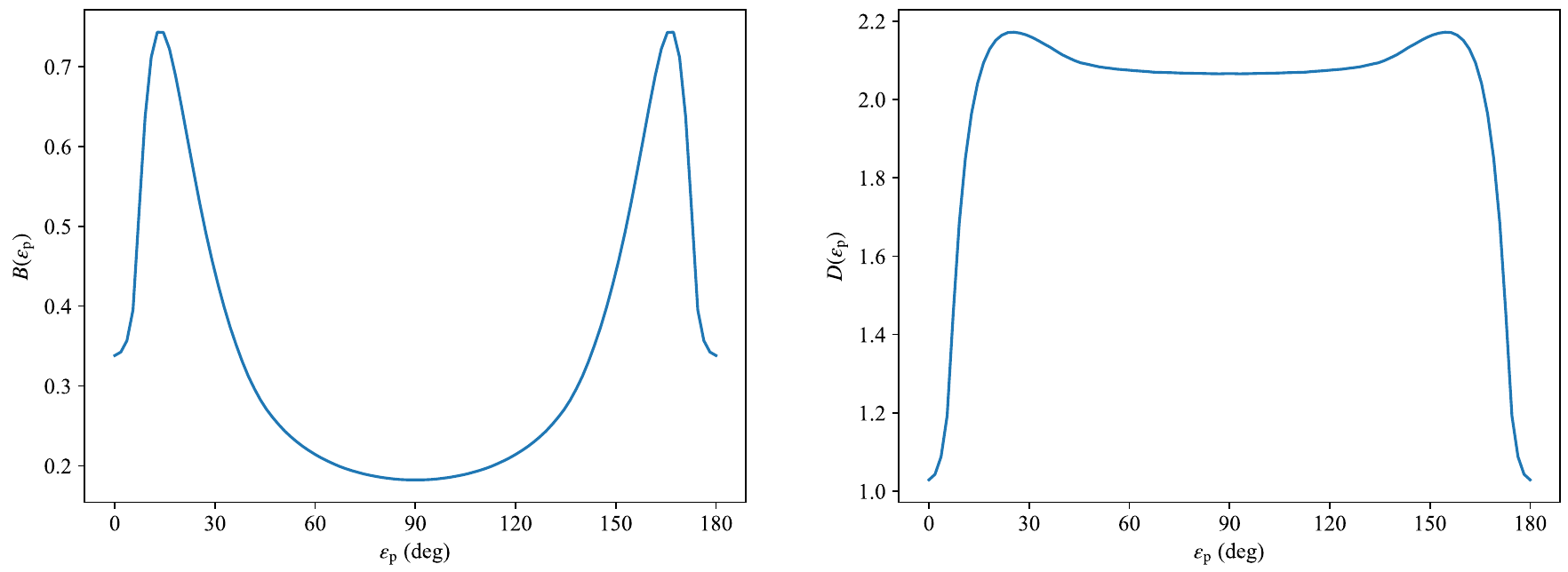}
    \caption{Values of $B(\varepsilon_{\rm p})$ and $D(\varepsilon_{\rm p})$ as a function of the planet obliquity $\varepsilon_{\rm p}$. The shadow fraction is $\eta_{\rm shadow}(\varepsilon_{\rm p}) = B(\varepsilon_{\rm p}) \cdot (R/r_{\rm p})^{D(\varepsilon_{\rm p})}$.}
    \label{fig:shadow_para}
\end{figure}

Then we fitted the numerically obtained shadow fractions with a power-law function of the form
\begin{equation}
    \eta_{\mathrm{shadow}}(\varepsilon_{\rm p}) = B(\varepsilon_{\rm p}) \left(\frac{R}{r_{\mathrm{p}}}\right)^{D(\varepsilon_{\rm p})}.
\end{equation}
For Saturn's case with $\varepsilon_{\rm p}=26.7^\circ$, the best-fit values were found to be $B \simeq 0.5$ and $D \simeq -2.17$, indicating that the fraction of time a ring particle remains in eclipse decays approximately as the inverse square of the orbital distance from the planet, as shown in Fig.~\ref{fig:eta_shadow}. The result is also valid for Martian system whose obliquity is $25.2^\circ$. Figure~\ref{fig:shadow_para} shows the values of $B(\varepsilon_{\rm p})$ and $D(\varepsilon_{\rm p})$ as a function of $\varepsilon_{\rm p}$ over the range (0,~180) degrees.

\section{Approximation for the EY coefficient}
\label{app:f_EY}

\begin{figure}
    \centering
    \includegraphics[width=0.5\linewidth]{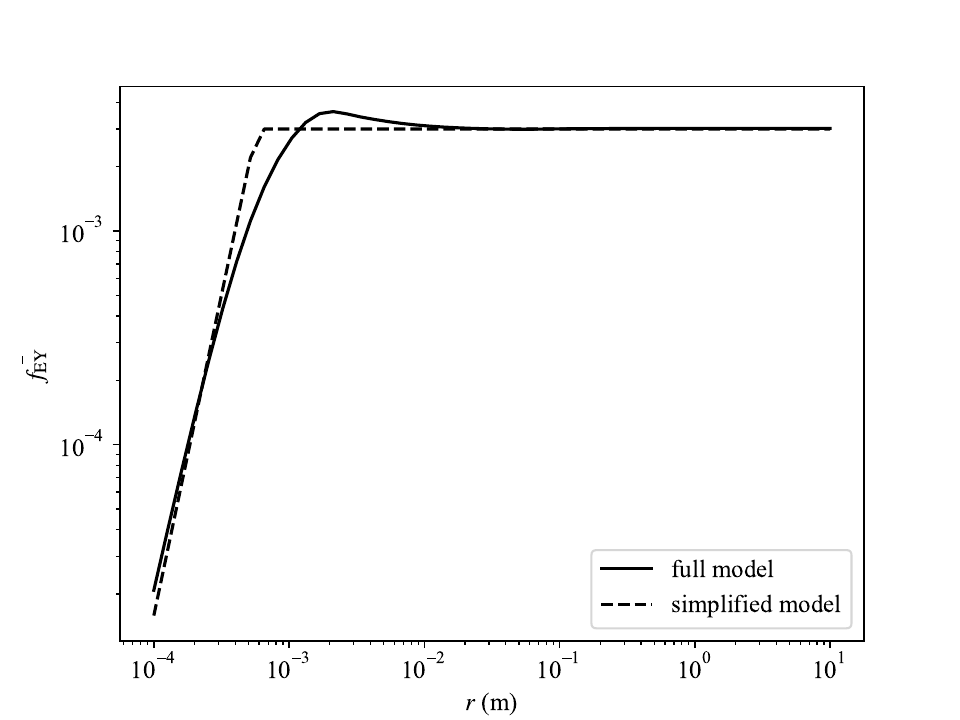}
    \caption{The EY coefficient dependence on the particle size. The full solution (Eqs.~\ref{eq:f_EY_d} and \ref{eq:f_EY_s}) can be approximated as a piecewise function (Eq.~\ref{eq:f_EY_size}) for easier integration.}
    \label{fig:f_EY_size}
\end{figure}

As stated in Sec.~\ref{sec:formula}, $f_\EY$ loses its dependence on the particle size when the size is larger than the characteristic thermal penetration $r_0$. When $r < r_0$, $f_\EY$ can be approximated to be proportional to $r^3$, as shown in the Fig.~\ref{fig:f_EY_size}.

\section{The formula of $\eta_{\rm size}$ considering a power law size distribution}
\label{app:eta_size}

\begin{figure}
    \centering
    \includegraphics[width=0.5\linewidth]{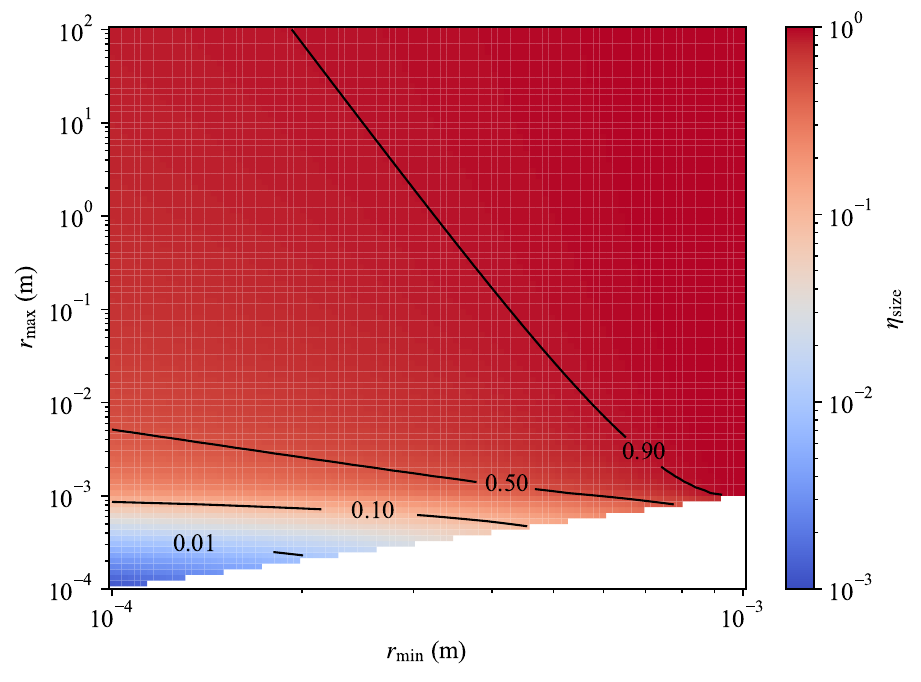}
    \caption{Size factor $\eta_{\rm size}$ as a function $r_{\rm min}$ and $r_{\rm max}$ for a power-law distribution with $\alpha = 3$. The $r_{\rm min}$ is set to range to $1~$mm since $\eta_{\rm size}$ equals 1 when $r_{\rm min} > 1~$mm. }
    \label{fig:f_size}
\end{figure}

The ring particles at an arbitrary layer with the optical depth $\tau'$ can be assumed to follow a power law ${\rm d}N = C r^{-\alpha} {\rm d}r$ with the minimum radius $r_{\rm min}$ and maximum radius $r_{\rm max}$. Constrained by the total mass, we obtain 
\begin{equation}
    C = {3 R \Delta R \Sigma (4-\alpha) \over 2\rho( r_{\max}^{4-\alpha} - r_{\rm min}^{4-\alpha})} {\Delta \tau' \over \tau}
\end{equation}
except for the singularity $\alpha = 4$ that leads to $C ={3 R \Delta R \Delta \tau'\Sigma / 2\rho \tau \ln (r_{\rm max}/r_{\rm min})} $. The total torque of all particles in the annulus by the Eclipse-Yarkovsky effect is 
\begin{equation}
    T_\EY = \int F_\EY R {\rm d}N = \int_{r_{\rm min}}^{r_{\rm max}} F_\EY R C r^{-\alpha} {\rm d}r.
\end{equation}
When $r_{\rm min}<r_0<r_{\rm max}$, by substituting $F_\EY$, we obtained
\begin{equation}
\label{eq:T_EY_size_1}
\begin{aligned}
    T_{\EY} &= {\pi (1 - A_{\rm v}) \,  \Phi_{\rm s} f_{\EY, 0} R C \over c r_0^3} e^{-\tau/\sin \psi}  {\Delta \tau' \over \tau} \left( \int_{r_{\rm min}}^{r_0} r^{5-\alpha} {\rm d}r + r_0^3 \int_{r_{0}}^{r_{\rm max}} r^{2-\alpha} {\rm d}r \right) \\
    &=  { \pi (1 - A_{\rm v}) \,  \Phi_{\rm s} f_{\EY, 0} R C \over c r_0^3} e^{-\tau/\sin \psi}  {\Delta \tau' \over \tau} \left( {r_0^{6-\alpha}-r_{\rm min}^{6-\alpha} \over 6-\alpha} + r_0^3 {r_0^{3-\alpha}-r_{\rm min}^{3-\alpha} \over 3-\alpha}   \right) \\
    &= {2\pi R^2 \Delta R (1 - A_{\rm v}) \,  \Phi_{\rm s} f_{\rm EY,0} \over c} {\Delta \tau' \over \tau} {3\Sigma \over 4 \rho r_0^3} {4-\alpha \over r_{\rm max}^{4-\alpha}  - r_{\rm  min}^{4-\alpha}} e^{-\tau/\sin \psi} \left( {r_0^{6-\alpha}-r_{\rm min}^{6-\alpha} \over 6-\alpha} + r_0^3 {r_0^{3-\alpha}-r_{\rm min}^{3-\alpha} \over 3-\alpha}   \right).
\end{aligned}
\end{equation}
Considering the definition of the optical depth
\begin{equation}
    \tau = {\sum_i \pi r_i^2 \over \sum m_i} {\Sigma} = {\Sigma } { \int_{r_{\rm min}}^{r_{\rm max}} \pi r^2 C r^{-\alpha} {\rm d} r \over \int_{r_{\rm min}}^{r_{\rm max}} 4\pi\rho r^3 C r^{-\alpha} {\rm d} r/3} = {3\Sigma \over 4 \rho} {(4 - \alpha)(r_{\rm max}^{3-\alpha} - r_{\rm min}^{3-\alpha} ) \over (3-\alpha) (r_{\rm max}^{4-\alpha} - r_{\rm min}^{4-\alpha} )},
\end{equation}
where $m_i$ is the mass of $i$-th particle, Eq.~\ref{eq:T_EY_size_1} becomes 
\begin{equation}
\label{eq:T_EY_size_2}
    T_{\EY} = {2\pi R^2 \Delta R (1 - A_{\rm v}) \,  \Phi_{\rm s} f_{\rm EY,0} \over c} {\Delta \tau' e^{-\tau/\sin \psi} } {3-\alpha \over r_{\rm max}^{3-\alpha}  - r_{\rm  min}^{3-\alpha}} \left( {r_0^{6-\alpha}-r_{\rm min}^{6-\alpha} \over r_0^3 (6-\alpha)} +  {r_{\rm max}^{3-\alpha}-r_{0}^{3-\alpha} \over 3-\alpha}   \right).
\end{equation}
Comparing Eq.~\ref{eq:T_EY_layer} with Eq.~\ref{eq:T_EY_size_2}, we obtain
\begin{equation}
\label{eq:f_size}
    \eta_{\rm size} =  {3-\alpha \over r_{\rm max}^{3-\alpha}  - r_{\rm  min}^{3-\alpha}} \left( {r_0^{6-\alpha}-r_{\rm min}^{6-\alpha} \over r_0^3 (6-\alpha)} +  {r_{\rm max}^{3-\alpha}-r_{0}^{3-\alpha} \over 3-\alpha}   \right).
\end{equation}

When $r_{\rm min} > r_0$, we have
\begin{equation}
    T_{\EY} = {\pi (1 - A_{\rm v}) \,  \Phi_{\rm s} f_{\EY, 0} R C \over c } e^{-\tau/\sin \psi}  {\Delta \tau' \over \tau}\left(  \int_{r_{\rm min}}^{r_{\rm max}} r^{2-\alpha} {\rm d}r \right).
\end{equation}
Following the same procedure as above, we can obtain
\begin{equation}
    \eta_{\rm size} =  1,
\end{equation}
which can also be reduced from Eq.~\ref{eq:f_size} by discarding the first term in the bracket and replace $r_0$ with $r_{\rm min}$ in the second term. The unity value of $\eta_{\rm size}$ is because $f_\EY$ has no dependence on the size $r$ when $r > r_0$.

After calculating the case when $r_{\rm max} < r_0$, we obtain the complete form of $\eta_{\rm size}$:
\begin{equation}
\label{eq:eta_size1}
\eta_{\rm size} =
\begin{cases}
 1, & r_{\rm max} > r_{\rm min} > r_0  \\
  {3-\alpha \over r_{\rm max}^{3-\alpha}  - r_{\rm  min}^{3-\alpha}} \left( {r_0^{6-\alpha}-r_{\rm min}^{6-\alpha} \over r_0^3 (6-\alpha)} +  {r_{\rm max}^{3-\alpha}-r_{0}^{3-\alpha} \over 3-\alpha}   \right), & r_{\rm max} > r_0 > r_{\rm min} \\
 {(3-\alpha)(r_{\rm max}^{6-\alpha} - r_{\rm min}^{6-\alpha}) \over r_0^3 (6-\alpha)(r_{\rm max}^{3-\alpha}  - r_{\rm  min}^{3-\alpha})} , & r_0 > r_{\rm max} > r_{\rm min} 
\end{cases}
\end{equation}

{For Saturn’s rings, observational studies suggest that the size–frequency distribution slope lies in the range $2.7 \lesssim \alpha \lesssim 3.2$ \citep{Miller2024}, which spans the singular point $\alpha = 3$ in $\eta_{\rm size}$ (the other singular point, $\alpha = 6$, is unlikely to be realized). In addition, theoretical modeling predicts $\alpha \simeq 3$ \citep{Brilliantov2015}. We therefore adopt $\alpha = 3$ in this work. In the case of $\alpha =3$, following the same procedure as described above, we have} 
\begin{equation}
\label{eq:eta_size2}
\eta_{\rm size} =
\begin{cases}
 1, & r_{\rm max} > r_{\rm min} > r_0  \\
 {1 \over \ln(r_{\rm max}/ r_{\rm  min})} \left( {r_0^{3}-r_{\rm min}^{3} \over 3 r_0^3 } +  \ln(r_{\rm max}/ r_{0})   \right), & r_{\rm max} > r_0 > r_{\rm min} \\
 {r_{\rm max}^{3} - r_{\rm min}^{3} \over 3 r_0^3 \ln (r_{\rm max}/ r_{\rm  min})}, & r_0 > r_{\rm max} > r_{\rm min} 
\end{cases}
\end{equation}
The distribution of $\eta_{\rm size}$ is plotted as a function $r_{\rm min}$ and $r_{\rm max}$ in Fig.~\ref{fig:f_size}.

{As inferred from Eq.~\ref{eq:eta_size1}, when $\alpha < 3$, which corresponds to a shallow SFD, the term $(r_{\rm max}^{3-\alpha} - r_{\rm min}^{3-\alpha}) \sim r_{\rm max}^{3-\alpha}$ unless $r_{\rm max}$ and $r_{\rm min}$ are close. This indicates the largest particles will dominate the $\eta_{\rm size}$. On the other hand, when $\alpha >3$, the term $(r_{\rm max}^{3-\alpha} - r_{\rm min}^{3-\alpha}) \sim -r_{\rm min}^{3-\alpha}$. Therefore the smallest particles dominate $\eta_{\rm size}$ in this case. When $\alpha = 3$, both the largest and smallest particles affect significantly $\eta_{\rm size}$, as inferred from Eq.~\ref{eq:eta_size2}.}

\section{Simulation parameters in \texttt{pkdgrav}}
\label{app:sim_para}

The parameters in the simulation generating Fig.~\ref{fig:EY_spin} is listed in Table~\ref{tab:sim_params}.

\begin{table*}[t]
\centering
\caption{Simulation parameters and mappings from code units to physical values.}
\label{tab:sim_params}
\begin{tabular}{@{} l l c c @{}}
\toprule
\textbf{Quantity} & \textbf{Symbol / Unit}  & \textbf{Physical value (from \#)} \\
\midrule
Number of particles                  & $N$                                                 & 1200 \\
Number of SSDEM steps                & $N_{\rm step}$                                 & 339{,}411 \\
SSDEM timestep                       & $\Delta t$                       & $1.2034~\mathrm{s}$ \\
Spring constant         & $k_n$                             & $54.8108~\mathrm{kg\,s^{-2}}$ \\
Central mass                         & $M_{\rm p}$                       & $5.7\times10^{26}~\mathrm{kg}$ \\
Orbital distance                     & $R$                              & $87,348~\mathrm{km}$ \\
Orbital frequency                    & $\Omega$                           & period $= 7.36946~\mathrm{h}$ \\
Patch width                          & $W$                               & $220.149~\mathrm{m}$ \\
Patch length                         & $L$                                & $220.149~\mathrm{m}$ \\
Patch thickness                      & $H$                               & $10.7333~\mathrm{m}$ \\
Mean particle radius                 & $\langle R\rangle$                & $1.07333~\mathrm{m}$ \\
Mean particle mass                   & $\langle m\rangle$                & $1.4477\times10^{4}~\mathrm{kg}$ \\
Mean Hill radius                     & $\langle r_{\rm H}\rangle$         & $2.50682~\langle R\rangle$ \\
Mean escape speed                    & $\langle v_{\rm esc}\rangle$       & $6.87457~\Omega\,\langle R\rangle$ \\
Optical depth                        & $\tau$                             & $0.1$ \\
Surface density                      & $\Sigma$                          & $358.669~\mathrm{kg\,m^{-2}}$ \\
Mean particle density                & $\rho_{\rm s}$                    & $2~\mathrm{g\,cm^{-3}}$ \\
Critical wavelength                  & $\lambda_{\rm crit}$             & $20.0082~\mathrm{m}$ \\
Friction angle &$\phi$   & $35^\circ$\\
Normal restitution coefficient  & $\epsilon_{\rm n}$ & 0.5 \\
Tangential restitution coefficient & $\epsilon_{\rm t}$ & 0.5 \\
\bottomrule
\end{tabular}
\end{table*}


\bibliography{references}{}
\bibliographystyle{aasjournal}



\end{document}